\documentclass[sigconf, screen,  nonacm]{acmart}

\usepackage{hyperref}
\usepackage{hyperxmp}

\usepackage{nicefrac}
\usepackage{siunitx}
\usepackage{array,framed}
\usepackage{booktabs}
\usepackage{colortbl}
\usepackage{svg}
\usepackage{xcolor}
\usepackage{enumitem}
\usepackage{
  color,
  float,
  epsfig,
  wrapfig,
  graphics,
  graphicx,
  subcaption
}
\usepackage{textcomp}
\usepackage{setspace}
\usepackage{latexsym,fancyhdr,url}
\usepackage{enumerate}
\usepackage{graphics}
\usepackage{xparse} 
\usepackage{xspace}
\usepackage{multirow}
\usepackage{csvsimple}
\usepackage{balance}
\usepackage{amsmath,amsfonts}

\usepackage{filecontents}

\usepackage{endnotes}
\usepackage{tcolorbox}
\usepackage{listings}
\usepackage{algorithm2e}
\RestyleAlgo{ruled}
\usepackage{algpseudocode}
\usepackage{tabularx} 
\usepackage{seqsplit}
\usepackage{soul}
\usepackage{fontawesome5}

\usepackage{
  tikz,
  pgfplots,
  pgfplotstable
}

\definecolor{bluegray}{rgb}{0.4, 0.6, 0.8}
\definecolor{darkgreen}{rgb}{0.1,0.5,0.1}
\definecolor{darkblue}{rgb}{0.3,0.3,0.8}

\newcommand{\eg}{\textit{e}.\textit{g}.,~}

\newcommand{\yicheng}[1]{\textcolor{blue}{[Yicheng: #1]}}

\newcommand{\nael}[1]{\textcolor{orange}{[Nael: #1]}}
\newcommand{\sankha}[1]{\textcolor{violet}{[Sankha: #1]}}

\renewcommand{\nael}[1]{}
\renewcommand{\yicheng}[1]{}
\renewcommand{\sankha}[1]{}

\newcolumntype{L}[1]{>{\raggedright\let\newline\\\arraybackslash\hspace{0pt}}m{#1}}
\newcolumntype{C}[1]{>{\centering\let\newline\\\arraybackslash\hspace{0pt}}m{#1}}
\newcolumntype{R}[1]{>{\raggedleft\let\newline\\\arraybackslash\hspace{0pt}}m{#1}}

\newcommand{\timeicon}{\faClock[regular]}

\newcommand{\timecounter}{\faClock[regular]~+~\faCalculator}

\begin{document}

\title{NVBleed: Covert and Side-Channel Attacks on NVIDIA Multi-GPU Interconnect}
\author{Yicheng Zhang}
\affiliation{%
  \institution{University of California, Riverside}
  \city{Riverside}
  \state{CA}
  \country{USA}
}
\email{yzhan846@ucr.edu}

\author{Ravan Nazaraliyev}
\affiliation{%
  \institution{University of California, Riverside}
  \city{Riverside}
  \state{CA}
  \country{USA}
}
\email{rnaza005@ucr.edu}

\author{Sankha Baran Dutta}
\affiliation{%
  \institution{Brookhaven National Laboratory}
  \city{Upton}
  \state{NY}
  \country{USA}
}
\email{sdutta2@bnl.gov}

\author{Andres Marquez}
\affiliation{%
  \institution{Pacific Northwest National Laboratory}
  \city{Richland}
  \state{WA}
  \country{USA}
}
\email{Andres.Marquez@pnnl.gov}

\author{Kevin Barker }
\affiliation{%
  \institution{Pacific Northwest National Laboratory}
  \city{Richland}
  \state{WA}
  \country{USA}
}
\email{kevin.barker@pnnl.gov}

\author{Nael Abu-Ghazaleh}
\affiliation{%
  \institution{University of California, Riverside}
  \city{Riverside}
  \state{CA}
  \country{USA}
}
\email{naelag@ucr.edu}


\pagestyle{plain} 

\begin{abstract}

Multi-GPU systems are becoming increasingly important in high-performance computing (HPC) and cloud infrastructure, providing acceleration for data-intensive applications, including machine learning workloads. These systems consist of multiple GPUs interconnected through high-speed networking links such as NVIDIA’s NVLink.
In this work, we explore whether the interconnect on such systems can offer a novel source of leakage, enabling new forms of covert and side-channel attacks.  Specifically,  we reverse-engineer the operations of NVlink and identify two primary sources of leakage: timing variations due to contention and accessible performance counters that disclose communication patterns.  The leakage is visible remotely and even across VM instances in the cloud, enabling potentially dangerous attacks.
Building on these observations, we develop two types of covert-channel attacks across two GPUs, achieving a bandwidth of over 70 Kbps with an error rate of 4.78\% for the contention channel.
We develop two end-to-end intra-VM side-channel attacks: application fingerprinting (including 18 high-performance computing and deep learning applications) and 3D graphics character identification within {\em Blender}, a multi-GPU rendering application. These attacks are highly effective, achieving F1 scores of up to 97.78\% and 91.56\%, respectively.
We also discover that leakage surprisingly occurs across Virtual Machines on the Google Cloud Platform (GCP) and demonstrate a side-channel attack on Blender, achieving F1 scores exceeding 88\%.
We also explore potential defenses such as managing access to counters and reducing the resolution of the clock to mitigate the two sources of leakage.

\end{abstract}

\maketitle

\section{Introduction}
\label{sec:intro}
Graphics Processing Units (GPUs) have emerged as a primary infrastructure supporting data-intensive applications. These applications range across a number of domains, including machine learning and natural language processing, scientific simulations, cryptocurrency mining, and 3D graphics rendering. As the size of these problems continues to increase,  multi-GPU computing is needed to match this expanding demand, which far surpasses the computational and memory resources of a single GPU. For example, training large language models (LLMs) is only possible through large numbers of GPUs: the LLaMA 65B-parameter model leveraged 2048 NVIDIA A100 GPUs across 21 days for its training~\cite{touvron2023llama}. Similarly, Smith et al.~\cite{smith2022using} trained the Megatron-Turing Natural Language Generation model (MT-NLG) on NVIDIA’s Selene supercomputer, utilizing 560 DGX A100 nodes (several thousand GPUs).   Despite the widespread adoption of GPUs across leading cloud computing platforms such as Amazon Elastic Compute Cloud (EC2), Microsoft Azure, and Google Compute Platform (GCP), handling sensitive private information, research into their security remains in its early stages.  

To support high-performance multi-GPU applications, multi-GPU systems use custom high-performance interconnects to support the bandwidth and latency requirements of these applications. For example, server-class NVIDIA GPUs leverage high-bandwidth interconnects such as NVLink and NVSwitch to achieve high throughput and low-latency communication. These links are used to communicate between the GPUs either explicitly using commands that copy data from one GPU to another or implicitly through transactions that access shared memory that is mapped to remote GPUs. 
 These communication patterns vary across applications and potentially with the sensitive data processed by applications: if an attacker is able to observe the communication patterns, they may be able to extract sensitive information about the applications and the data they are processing.

 A number of previous studies~\cite{naghibijouybari2017constructing, frigo2018grand, naghibijouybari2018rendered, wei2020leaky,  ahn2021network, side2022lockeddown,  zhang2023t, zhang2024invalidate+} presented covert and side-channel vulnerabilities of a single GPU, or between a Central Processing Unit (CPU) and GPU~\cite{dutta2021leaky}.  
One prior work by Dutta et al.~\cite{dutta2023spy} proposed a microarchitectural side-channel attack on multi-GPU caches.  Our work presents an orthogonal source of leakage that leaks the communication behavior through the multi-GPU interconnect, leaking cross-GPU communication rather than memory access patterns.  Our attack also offers advantages in terms of attacker co-location.  As a result, the attack substantially expands our understanding of the side-channel threats faced by these systems and the types of defenses needed to mitigate them.  Although some prior works have targeted interconnects within the processor chip~\cite{dutta2021leaky,wan2022meshup,paccagnella2021lord} or the bus interconnect between a CPU and peripherals~\cite{tan2021invisible,side2022lockeddown}, to our knowledge, this is the first attack to exploit multi-GPU communication.


 
 We reverse-engineer the NVLink behavior and characterize the available sources of leakage that an attacker may exploit. Our investigation reveals that NVIDIA provides user-level accessible performance counters related to NVLink transactions that leak information across applications and even across instances in the real cloud.
 In addition, we observe timing differences in the NVLink transactions that leak information about other ongoing communication.  
 Through monitoring these readings, we managed to reverse-engineer the NVLink packet format and communication patterns between multiple GPUs (discussed in §~\ref{sec:nvlink_analysis}).  We leverage this leakage to build both covert and side-channel attacks. 
 
We demonstrate attacks on two generations of multi-GPU systems: a local server (DGX-1 system with Tesla P100 GPU) and a public cloud instance (Google Compute Platform with 8 Tesla V100 GPUs), showing the general nature of these attacks. 
First, we develop two covert-channel attacks ({\em ContenLink} and  {\em LeakyCounterLink}) in which a sender program on one GPU transmits information covertly to a receiver program on another GPU (discussed in §~\ref{sec:covert_channel}). We optimize the channel by incorporating additional parallelism, attaining a bandwidth of 70.59 Kbps, with an error rate of 4.78\% for the contention channel. 

We also carry out two end-to-end intra-VM side-channel attacks. In the first attack, we demonstrate that by probing the NVLink performance counters, an attacker can infer the activities of a concurrent application, allowing them to fingerprint applications such as deep learning models and physical dynamics simulation benchmarks (discussed in §~\ref{subsec:app_finger}), with 97.8\% accuracy. The second attack demonstrates that we can recover data-dependent leakage, identifying which 3D graphics character is being rendered by a victim user running the popular Blender rendering toolkit~\cite{blender} (discussed in §~\ref{subsec:3d_finger}), with an accuracy exceeding 91\%. We investigated the leakage across different virtual machine instances, which use different GPUs, and discovered, surprisingly, that there is measurable leakage even though the VMs do not communicate.  We leverage this leakage to develop a cross-VM side-channel attack on the public cloud platform GCP, achieving an F1 score of over 88\% in correctly identifying 3D rendered characters (discussed in §~\ref{sec:cross-instance}). 
Finally, we discuss the challenges associated with mitigating these attacks and propose potential solutions in §~\ref{sec:mitigation}.

In summary, the contributions of this paper are:
\begin{itemize}
\setlength{\itemsep}{0in}
    \item We reverse-engineer the NVLink interconnect and identify two leakage sources disclosing aspects of their operation.
    \item We introduce two types of intra-VM covert-channel attacks that exploit timing and performance counters on two different multi-GPU systems.
    \item We demonstrate two NVLink-based intra-VM side-channel attacks: (1) fingerprinting applications and (2) identifying 3D graphics character rendering on the victim GPU.
    \item We identify NVLink leakage across two co-located VM instances and develop a cross-VM side-channel attack capable of identifying 3D graphics characters.


\end{itemize}




\section{Background}
\label{sec:Background}

In this section, we provide background on NVLink and an overview of how the GPU performance monitoring unit operates.


\subsection{NVLink}
\label{subsec:nvlink_bg}
NVLink is a high-speed, high-bandwidth interconnect technology developed by NVIDIA~\cite{nvlink_whitepaper} to support data sharing among peer GPUs~\cite{li2019evaluating}, as well as between CPUs and GPUs.  NVLinks are bidirectional, comprising two sublinks, one for each direction.  NVLink supports communication between the GPUs in three different ways: (1) Explicit Peer-to-Peer (P2P) communication enables GPUs to copy data using \texttt{cudaMemcpyPeer()}, which copies a user-specified size buffer across the GPUs; (2) Pinned (device) memory, allocated via \texttt{cudaMalloc()}, can be accessed by other GPUs without copying it to the host; and (3) Unified virtual memory: in this mode, GPUs are able to access memory that is mapped on remote GPUs.  When this happens, either data is migrated at the page boundary or returned at a cache line granularity; these decisions are determined by programmer hints as well as run-time considerations. This permits efficient reading and writing on the remote CPU's host memory and the device memory of a peer GPU. 

\begin{figure}[tbh]
    \begin{subfigure}[b]{0.3\textwidth}
    \includegraphics[width=\textwidth]{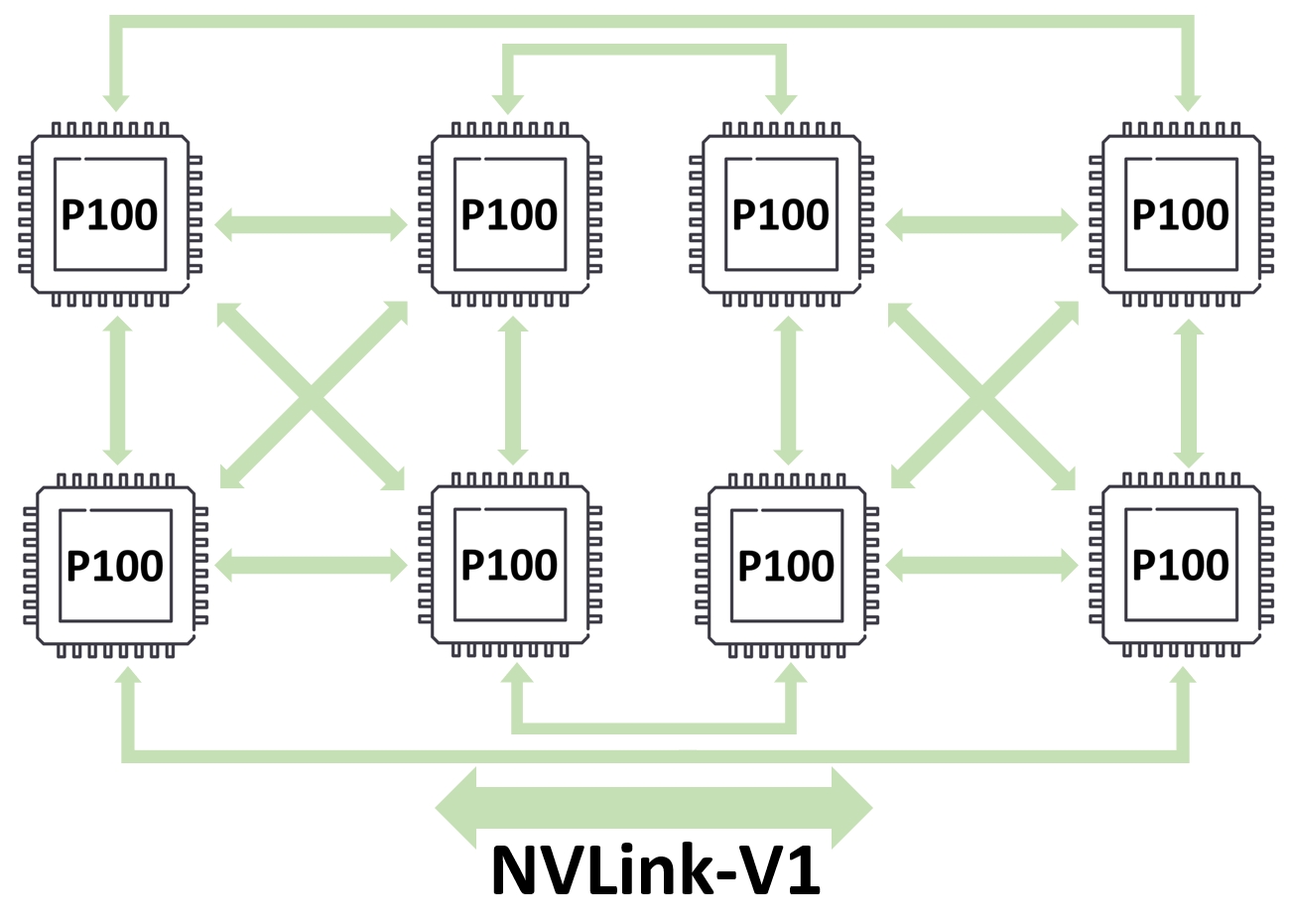}
    \caption{DGX-1 system -- P100 GPUs with 4 NVLinks.}
    \label{fig:dgx_topo}
    \end{subfigure}
    \hfill 
    \begin{subfigure}[b]{0.3\textwidth}
    \includegraphics[width=\textwidth]{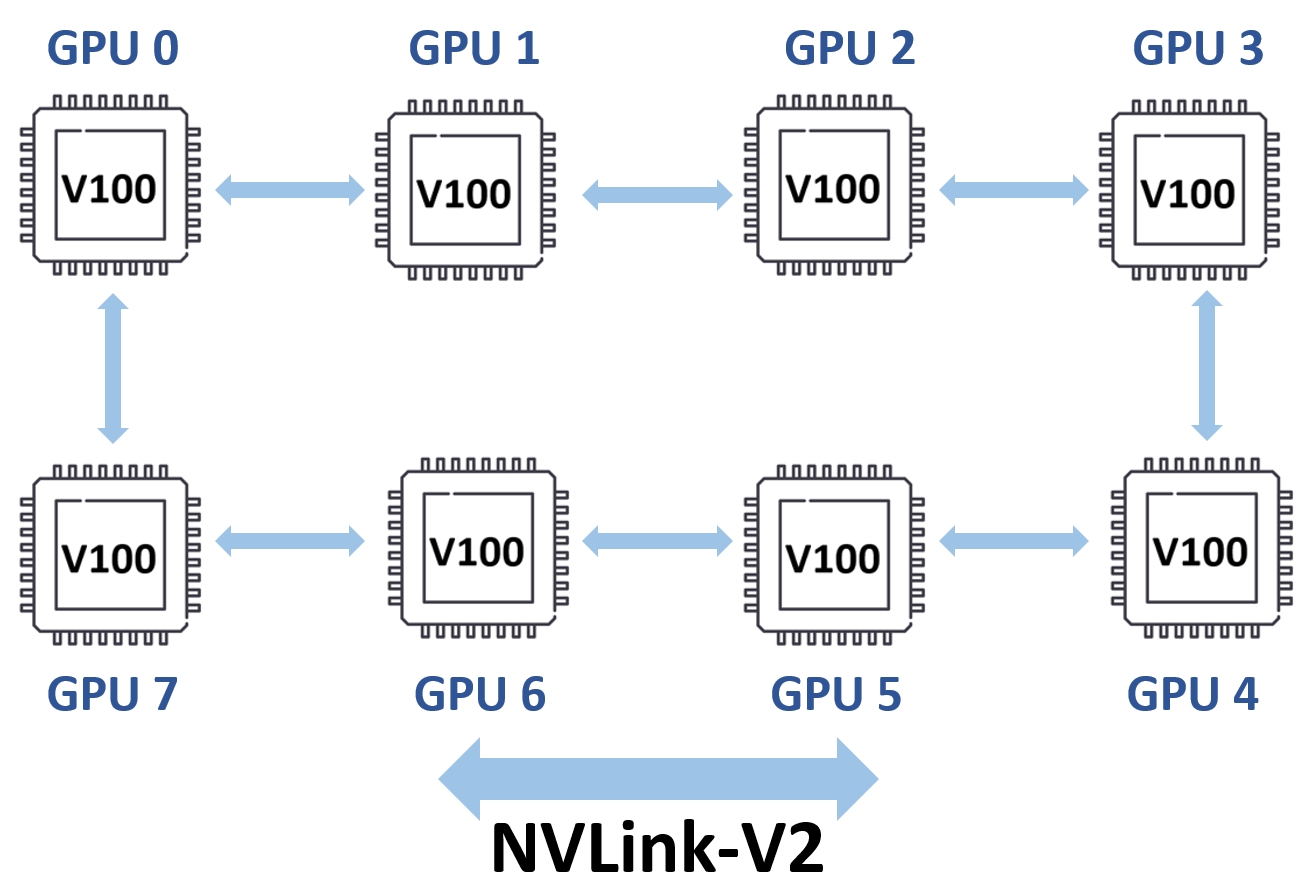}
    \caption{GCP 8-GPU machine -- V100 GPUs with 6 NVLinks.}
    \label{fig:gcp_topo}
    \end{subfigure}
    \caption{GPU topology of two experimental machines.}
    \label{fig:topo_overview}
\end{figure}

NVLink progressed through several generations: NVLink-V1 to V3 
and NVSwitch. NVLink-V1 was introduced with NVIDIA's Pascal P100 GPU~\cite{danskin2016pascal},  featuring 4 NVLink slots per GPU, each achieving a bandwidth of 20 GB/s. NVLink-V2, released with the Volta V100 GPU~\cite{choquette2018volta}, includes 6 NVLink slots, each with a bandwidth of 25 GB/s. Fig.~\ref{fig:topo_overview} illustrates the GPU topology of these NVLink generations. NVLink-V3 and NVSwitch are featured in the A100 GPU~\cite{choquette2021nvidia}, which is equipped with 12 NVLink links, each with a bandwidth of 50 GB/s. 
Recently, other GPU manufacturers, excluding NVIDIA, such as AMD, Intel, and others, are beginning to develop a comparable multi-GPU interconnect, the Ultra Accelerator Link (UALink), to compete with NVIDIA's NVLink~\cite{ualink_news}. 
In this study, we conducted experiments encompassing the two generations of NVLink (NVLink-V1 and V2). We leave NVLink-V3 and NVSwitch for future work.  

\subsection{GPU Performance Counter}
\label{subsec:gpu_per_ctr_bg}


GPU vendors have introduced performance monitoring units, similar to those available on CPUs, to help developers understand and optimize application performance. While these tools provide valuable insights, previous works~\cite{naghibijouybari2018rendered, wei2020leaky, hu2020deepsniffer, wang2022demystifying, patwari2022dnn, side2022lockeddown, taneja2023hot,yangSmartPhoneGPU} have shown they can be sources of side-channel leakage. 
The NVIDIA GPU performance counters are accessed through the CUDA Profiling Tools Interface (CUPTI)~\cite{CUPTI}.  
There are a number of NVLink-related performance counters; for instance, \emph{nvlink\_total\_da\-ta\_received} tracks the number of bytes of data received across all NVLinks on a particular GPU. AMD also offers analogous APIs for profiling GPU memory and communication transaction metrics~\cite{amdprof}. We comprehensively overview all available NVLink-related performance counters in Section~\ref{subsec:available_leakage_vector}.\nael{Remove this sentence if we are not keeping the table} \yicheng{It would be better to keep the table.}

\section{Threat Model and Attack Overview}
In this section, we present our threat model for two attack scenarios based on the placement of the spy and victim. We then summarize the attacker’s capabilities in each scenario and provide an overview of all relevant attacks.

\subsection{Threat Model}
\label{sec:threatmodel}

\begin{figure}[tbh]
    \centering
    \includegraphics[width=0.45\textwidth]{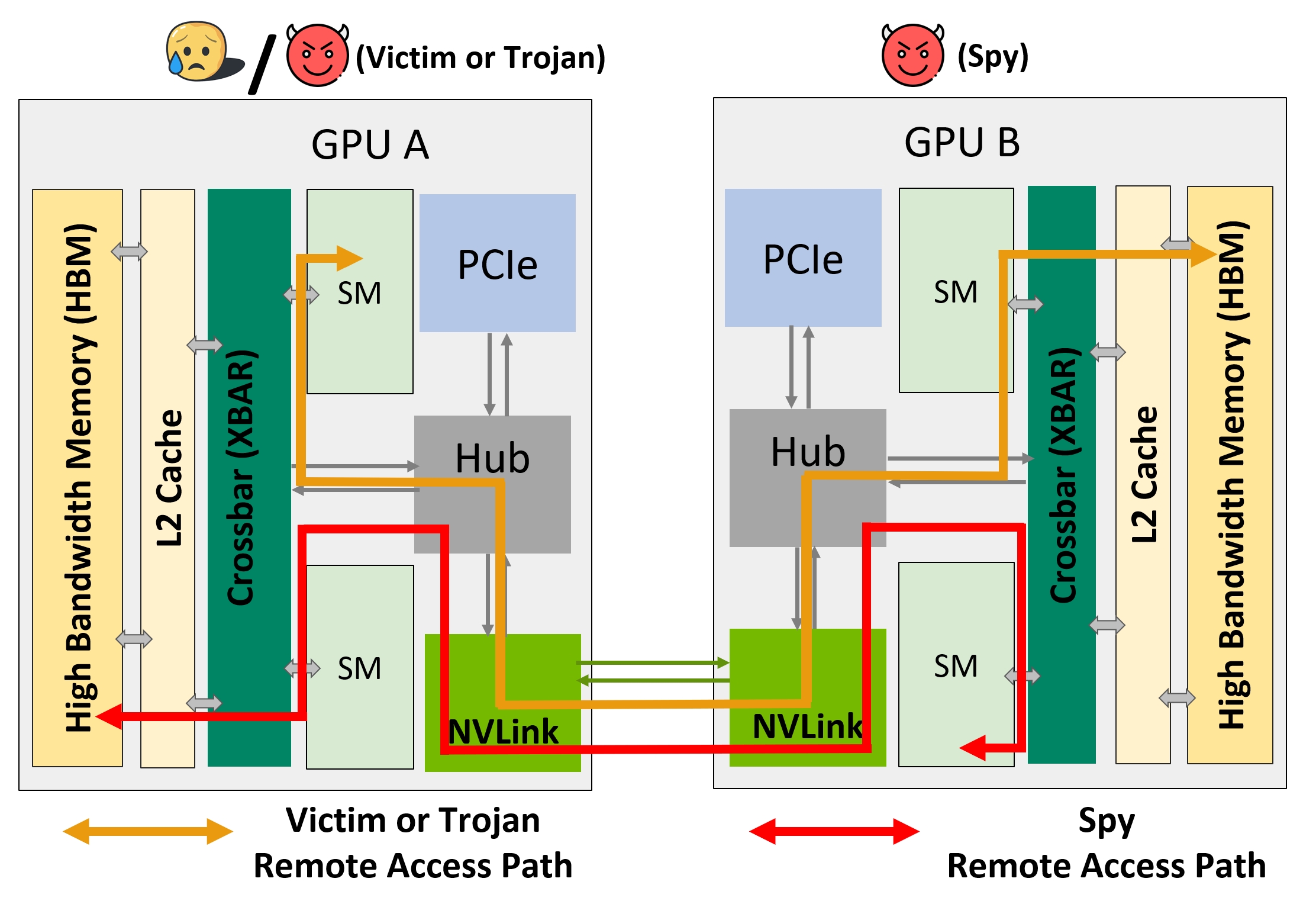}
    \caption{Intr-VM covert and side-channel attacks.}
    \label{fig:threat_model}
\end{figure}



We consider two attack scenarios: (1) Intra-VM attack, where multiple users are co-located on the same GPU system (e.g., NVIDIA DGX) or VM instance on the cloud and share NVLink connections, and (2) Cross-VM attack, where each VM has its own GPU(s), but the GPUs remain interconnected via an unused NVLink.

\noindent \textbf{Intra-VM attack.} In this scenario, the attacker and victim reside on different GPUs but share the same NVLink. Fig.{~\ref{fig:threat_model}} provides an overview of this setup. The attacker can exploit two capabilities:

\begin{itemize}[leftmargin=0.1in]
    \item \textbf{Measuring NVLink contention timing}: Without superuser privileges, the attacker can introduce contention on NVLink to observe timing differences, thereby inferring the victim’s behavior.
    \item \textbf{Accessing NVlink performance counters}:  NVIDIA supports performance counters that show traffic information on the shared NVLink connections.  However, NVIDIA released a driver patch{~\cite{nvidiapatch}} that optionally restricts performance counter access in user mode, so this leakage vector is only available if this patch is not applied or if the attacker can ask for downgraded drivers to be able to use performance counters.  Since performance counters are not reliably available in this threat model, we rely primarily on contention data (the first source of leakage) to estimate the counter values. For example, the counter {\texttt{nvlink\_receive\_thr\-oughput}} can be approximated by continuously transferring a fixed data size and measuring the transfer time:
    \begin{equation}
    Throughput = \frac{NVLink\_transmission\_size}{Data\_transmission\_time} \label{equation:synthetic}
    \end{equation}

\end{itemize}

\noindent \textbf{Cross-VM attack.} In this scenario, the attacker and victim run on different VMs; however, both VMs’ GPUs share an NVLink connection (see Fig.{~\ref{fig:cross_instance}}). Due to VM isolation, the attacker cannot directly cause contention with the victim, so there is no timing-based leakage. Nonetheless, major cloud providers (e.g., AWS, GCP, Azure, Alibaba) allow users to install any version of GPU drivers. Consequently, the attacker can downgrade her VM’s driver to enable performance counter collection, while the victim does not need to downgrade. By doing so, the attacker can infer the victim’s NVLink transactions, even across VMs.

\subsection{Overview of Attacks}
\label{subsec:attack_overview}

\textbf{Intra-VM covert channel attack.} We introduce two types of intra-VM covert channel attacks: \textit{ContenLink}  and \textit{LeakyCounterLink}. We assume the sender and receiver processes run on separate GPUs in both cases. In \textit{ContenLink}, we exploit NVLink contention leakage to establish a covert channel between two GPUs. In \textit{LeakyCounterLink}, we exploit NVLink leaky counters to demonstrate a covert channel attack (please note the potential access limitations to this leakage described above).  We evaluate the effectiveness of these attacks using two metrics: bandwidth and error rate.

\noindent 
\textbf{Intra-VM side-channel attack.} Next, we present two end-to-end intra-VM side-channel attacks. In both attacks, the victim and attacker processes run on two different GPUs but share the same NVLink. The first demonstrates application fingerprinting across 18 different applications. The second showcases 3D character fingerprinting using 50 characters from Blender Studio{~\cite{blender_character}}. To assess the performance of these classifiers, we calculate three metrics: F1 score (F1), Precision (Prec), and Recall (Rec). 

\noindent 
\textbf{Cross-VM side-channel attack.} Finally, we demonstrate an end-to-end cross-VM side-channel attack. We assume the attacker and victim reside in separate VMs connected via an unused NVLink. Here, the attacker aims to identify which 3D characters are being rendered in the victim’s VM. The same feature extraction and evaluation metrics from our earlier side-channel attacks are applied in this scenario as well.

\section{Demystifying NVLinks}
\label{sec:nvlink_analysis}

In this section, we first reverse-engineer the NVLink operation. Next, we analyze the NVLink-related performance counters and identify their leakage. Finally, we study the timing behavior when NVLink contention arises to identify timing-related leakage.


\noindent \textbf{{Experiment platform.}} We use a local DGX-1 multi-GPU system as well as a public cloud server from Google as our experimental environments (details in Table~\ref{tb:experiment_platform}). Fig.~\ref{fig:topo_overview} presents an abstract view of the NVLink topology in these systems. The DGX-1 system employs a hypercube topology, while the GCP 8-GPU platform uses a ring topology. 
We tested the GCP VMs in three different regions: us-central1-c, us-east1-c, and us-west1-a.

\begin{table}[htb]
\caption{Specification of two target platforms.}
\small
\label{tb}
\begin{tabular}{l|l|l|}

\cline{2-3}
\textbf{}                        & \textbf{DGX-1 system}           & \textbf{Server (GCP)} \\ \hline \hline
\multicolumn{1}{|l|}{CPU}        & Intel Xeon E5-2698v4 & GCP N1-standard     \\ \hline
\multicolumn{1}{|l|}{Memory}     & 256GB                           & 100 GB                \\ \hline
\multicolumn{1}{|l|}{GPU}        & 8 Tesla P100s                   &  8 Tesla V100s                     \\ \hline
\multicolumn{1}{|l|}{OS}         & Ubuntu 22.04                    & Ubuntu 20.04          \\ \hline
\multicolumn{1}{|l|}{CUDA}       & V12.2                    & V10.1.243             \\ \hline
\multicolumn{1}{|l|}{GPU driver} & 535.129.03                        &        525.125.06               \\ \hline
\multicolumn{1}{|l|}{NVLink version} & NVLink-V1                      &    NVLink-V2               \\ \hline
\multicolumn{1}{|l|}{NVLink slots} & 1 for peer GPUs                      &    3 for peer GPUs               \\ \hline

\end{tabular}
\label{tb:experiment_platform}
\end{table}

\subsection{Reverse Engineering NVLink Operation}
\label{subsec:re_nvlink}

\begin{figure}[tbh]
    \centering
    \includegraphics[width=0.4\textwidth]{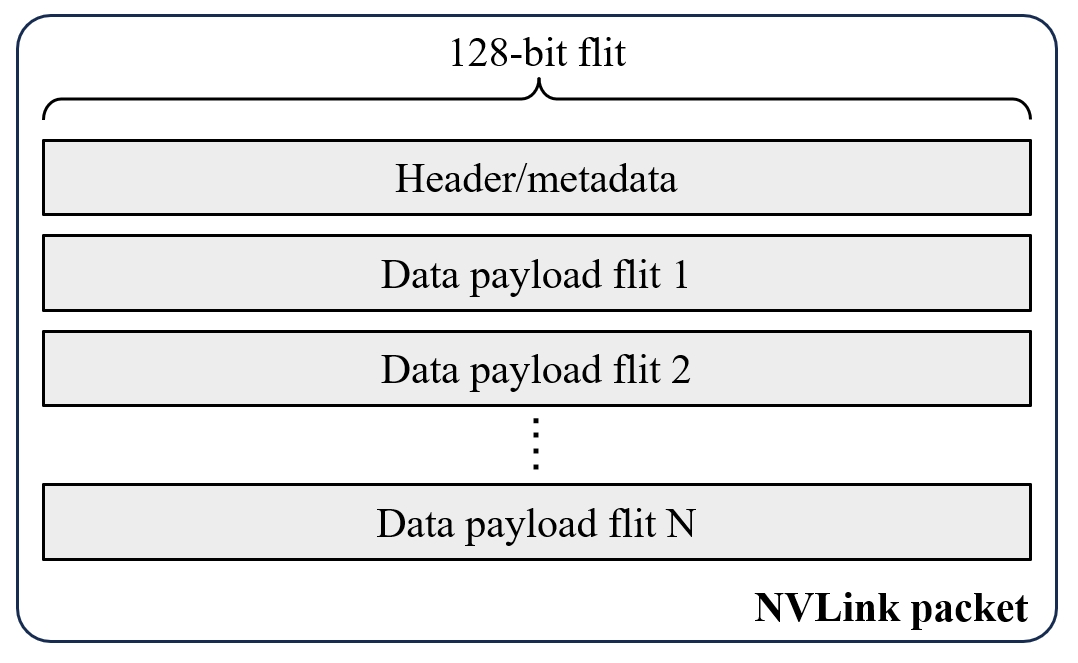}
    \caption{NVLink-V1 packet format.}
    \label{fig:nvlink_packet}
\end{figure}

NVLink operates as a packet-based interface, where each packet can contain multiple Flow Control Units (flits). Fig.~\ref{fig:nvlink_packet} illustrates the format of an NVLink packet. In this section, we use NVLink-related counters to reverse-engineer NVLink characteristics, such as the packet format.

\begin{figure*}[tbh]
    \centering
    \includegraphics[width=1\textwidth]{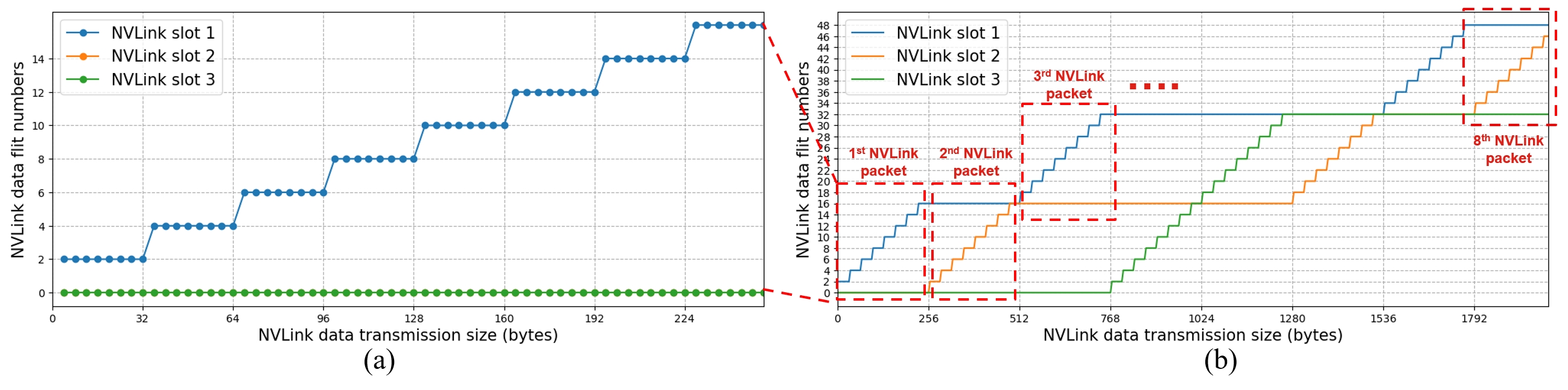}
    \caption{Reverse engineering NVLink transmission (NVLink-V2): (a) First packet, sizes 0--256 bytes, (b) Multiple packets.} 
    \label{fig:nvlink_re}
\end{figure*}

\noindent \textbf{{NVLink-V1 (DGX-1 system).}}  As discussed in an NVIDIA whitepaper~\cite{foley2017ultra}, NVLink-V1 utilizes a uniform packet format comprising a header, metadata, and $N=16$ data payload flits. Each flit is 128 bits (16 bytes), enabling the transfer of $16 \times N = 256$ bytes of data per NVLink packet. The minimum data transmission size on NVLink-V1 is 32 bytes, corresponding to 2 data payload flits.

\noindent \textbf{{NVLink-V2 (GCP V100).}} However, the packet format for NVLink-V2 has not been documented. Therefore, we design experiments to reverse-engineer this format. Specifically, we initiate data transfers from GPU 1 to GPU 0 via NVLink using \texttt{cudaMemcpyPeer()}, varying the size of the data transferred.
Concurrently, we profile the counter \texttt{nvlink\_user\_data\_rec\-eived} on GPU 0. This counter is chosen because it excludes headers and metadata, accurately monitoring the NVLink data transmission size.

Fig.~\ref{fig:nvlink_re} illustrates the usage of data flits with an increasing number of data transmissions on NVLink-V2 (GCP V100 machine). We obtain 3 separate readings as GPUs 0 and 1 in GCP machines are equipped with 3 NVLink slots. As depicted in Fig.~\ref{fig:nvlink_re} (b), we observe 8 distinct steps for every 256 bytes. We notice that the 3 NVLink slots are sequentially activated as the data transmission size increases. For example, when the data transmission size is less than 256 bytes, only slot 1 (indicated in blue) is activated, while the other two slots remain idle. However, when the data size exceeds 256 bytes, slot 2 (indicated in yellow) is also engaged to handle the additional data transfer. Thus, this observation leads us to conclude that the maximum packet size for NVLink-V2, consistent with NVLink-V1, is 256 bytes, comprising 16 data payload flits.

Furthermore, Fig.\ref{fig:nvlink_re} (a) depicts a zoomed-in view of the first NVLink packet. In the first NVLink packet, we observe 8 steps for every 32 bytes, corresponding to the size of 2 data flits, indicating that the size of the packet is increased at a granularity of 2 flits. For example, we note that slot 1 (indicated in blue) uses 2 data flits (32 bytes), even when the data transfer size is less than 32 bytes, which appears to be the smallest transaction size. 

\begin{tcolorbox}[colback=gray!10,colframe=black,boxrule=0.5pt,arc=2mm,outer arc=2mm,]
\textit{\textbf{Observation 1}:  The minimum data transmission size for both NVLink-V1 and NVLink-V2 is 32 bytes, equivalent to 2 data payload flits, and increases with a two-flit granularity. Each NVLink-V1/V2 packet contains a maximum of 16 data payload flits, totaling 256 bytes.  Transmissions use the available sublinks concurrently when possible although the pattern is not consistent.} 
\end{tcolorbox}

\begin{table*}[tbh]
\centering
\small
\caption{NVLink-related performance counters, available from NVIDIA CUPTI~\cite{CUPTI}.}
\vspace{-0.1in}
\begin{tabular}{|L{2.3cm}|L{15cm}|}

\hline
\textbf{Category} & \textbf{Counter Name}                                                                                                       \\ \hline \hline
Throughput         & nvlink\_receive/transmit\_throughput 
 \\ \hline
User              & nvlink\_user\_data\_received/transmitted, nvlink\_user\_write\_data\_transmitted, nvlink\_user\_response\_data\_received    \\ \hline
Total             & nvlink\_total\_data\_received/transmitted, nvlink\_total\_response\_data\_received, nvlink\_total\_write\_data\_transmitted                                                                                       \\ \hline
Atomic operation  & nvlink\_total/user\_nratom\_data\_transmitted, nvlink\_total/user\_ratom\_data\_transmitted                                 \\ \hline
\end{tabular}
\label{tb:allcounters}

\end{table*}

\subsection{NVLink Performance Counters}
\label{subsec:available_leakage_vector}


We collect all NVLink-related counters from CUPTI and evaluate them for potential leakage: Table~\ref{tb:allcounters} summarizes the 14 NVLink-related counters.

\noindent \textbf{{Experiment 1: Receive vs transmit.}} All NVLink counters feature both \emph{\texttt{receive}} and \emph{\texttt{transmit}} attributes. According to NVIDIA's white paper~\cite{foley2017ultra}, an NVLink transaction begins with a request from the requester GPU, followed by the target GPU transmitting the data payloads back to the requester. The request consists only of metadata describing the requested data, making it significantly smaller than the payloads transmitted from the target GPU.
We profile all NVLink counters by transmitting varying-sized packets using \texttt{cudaMemcpyPeer()}. We observe that on GPU 0 (receiver GPU), the \emph{\texttt{receive}} counters consistently show significantly higher values than the \emph{\texttt{transmit}} counters. 
Therefore, by examining NVLink receive/transmit values, the attacker can determine the victim’s data transfer direction.


\begin{tcolorbox}[colback=gray!10,colframe=black,boxrule=0.5pt,arc=2mm,outer arc=2mm,]
\textit{\textbf{Observation 2}: The NVLink \textbf{receive/transmit} attributes reveal NVLink data transaction direction.}
\end{tcolorbox}

\noindent \textbf{{Experiment 2: User vs total.}} We observed that, aside from throughput, all categories feature both \emph{\texttt{user}} and \emph{\texttt{total}} attributes; these attributes are undocumented, so we explore them using the following experiment. 

\begin{figure}[tbh]
    \centering
    \includegraphics[width=0.45\textwidth]{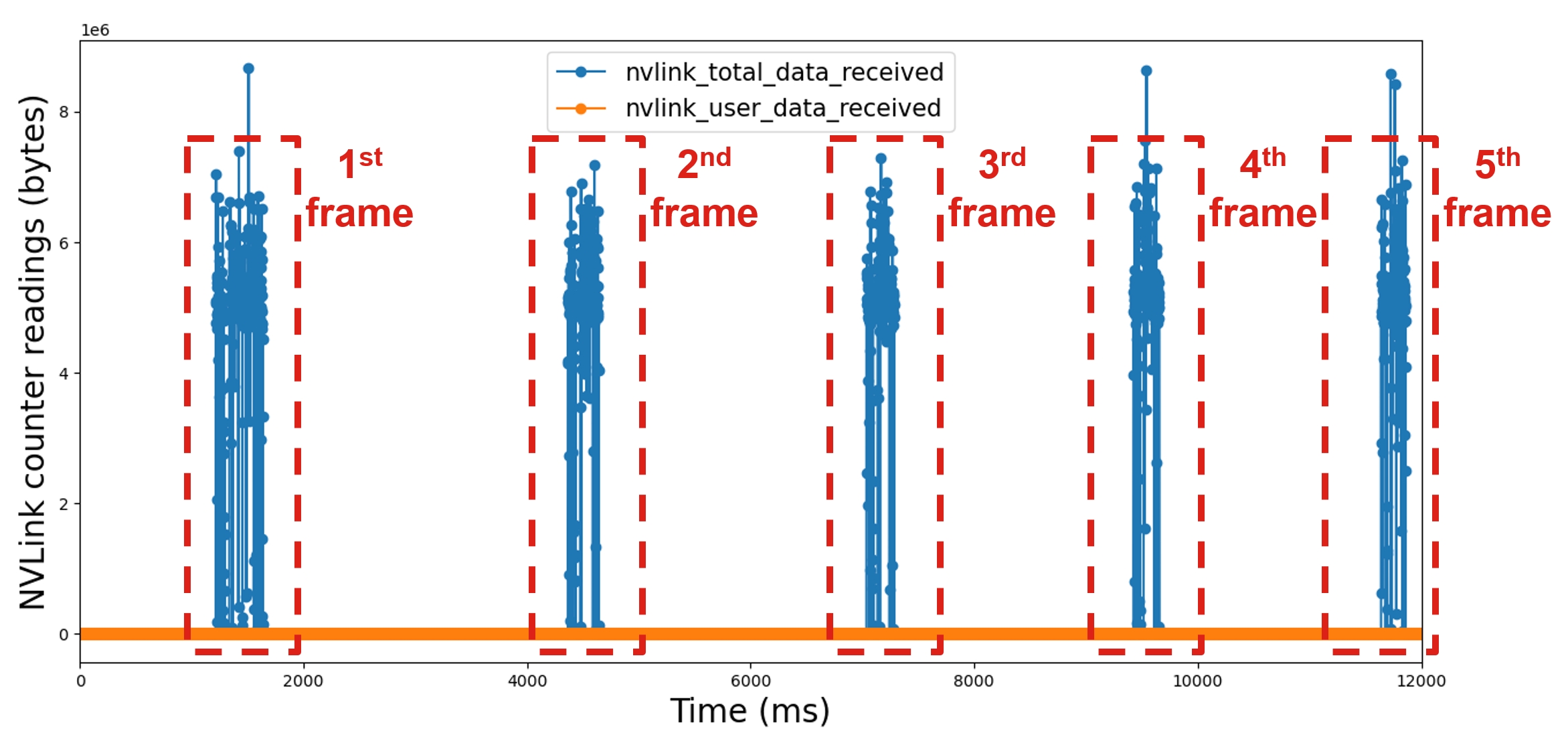}
    \caption{Counter traces for five consecutive frames.}
    \label{fig:blender_5_frame}
\end{figure}

In this experiment, a spy program continuously transfers data from GPU 0 to GPU 1 while profiling NVLink's \emph{\texttt{user}} and \emph{\texttt{total}} counters. Concurrently, we execute a program with a known NVLink communication pattern and observe the performance counters; in this experiment, we use a 3D rendering tool, Blender, to render 5 frames. Blender transfers 3D object data in each frame using the shared NVLink.  As Fig.~\ref{fig:blender_5_frame} shows, \emph{\texttt{user}} counters remain unaffected. However, the \emph{\texttt{total}} counters aggregate the transaction data from all programs using a shared NVLink. The 5 NVLink transmissions from Blender are clearly visible to the spy program.  
Therefore, by measuring the total NVLink values, the attacker can estimate the victim’s data transmission size.

\begin{tcolorbox}[colback=gray!10,colframe=black,boxrule=0.5pt,arc=2mm,outer arc=2mm,]
\textit{\textbf{Observation 3}: When NVLink is shared, the NVLink \textbf{total} counters reveal all data transaction patterns, including those from other users. }
\end{tcolorbox}




\noindent \textbf{{Experiment 3: Aggregation mode.}} As illustrated in Fig.~\ref{fig:topo_overview}, P100 GPU (NVLink-V1) and V100 GPU (NVLink-V2) are equipped with 4 and 6 NVLink slots, respectively.
NVLink offers an aggregate mode for performance counters that may be disabled~\cite{nvprof}. We discovered that when aggregation mode is turned off, the counters detail the data transaction sizes for each individual NVLink slot. In the DGX-1 system, we obtain 4 values per GPU, each corresponding to one link. In contrast, for the GCP setup, we receive 6 values per GPU. 


\begin{tcolorbox}[colback=gray!10,colframe=black,boxrule=0.5pt,arc=2mm,outer arc=2mm,]
\textit{\textbf{Observation 4}:  When aggregation mode is off, counter values are available for individual NVLink slots, providing finer-grain leakage.}
\end{tcolorbox}


\noindent \textbf{{NVLink counter selection.}} 
As a result of these experiments, we identify two counters, \texttt{nvlink\_total\_data\_received} and \texttt{nvl\-ink\_total\_data\_transmitted}, to use in our attacks.
We rule out user counters since they do not leak information, while total counters leak transmitted data information regarding other applications. 
 In addition, we observed that \emph{\texttt{throughput}} counters incur higher overhead, limiting the rate of obtaining information. Finally, the counters tracking atomic operations remain zero for applications that do not use remote atomic operations, limiting their utility.





\subsection{Timing-Based Leakage Due to Contention}
\label{subsec:contention_nvlink}
The final set of reverse engineering experiments targets timing properties under contention.  Contention occurs when multiple components simultaneously access shared resources, such as buses or I/O ports, which are constrained by bandwidth or capacity limits. 
In our threat model (Fig.~\ref{fig:threat_model}), the victim program is on GPU A, and the spy program is on GPU B. 
The victim accesses data from GPU B via NVLink while the spy fetches data from GPU A, leading to contention on the Crossbar (XBAR), High-speed Hub (HSHUB), and NVLink I/O ports. The XBAR enables data exchange between GPU SM cores, L2 cache, and high bandwidth memory (HBM)~\cite{xbar}. The HSHUB connects the XBAR to the GPU's I/O ports (PCIe or NVLink)~\cite{p100}. As they contend for these resources, this concurrent use of the NVLink from both programs results in observable delays compared to when the link is idle.


We use two programs in our experiments: (1) Program \textbf{A}, which continuously measures execution time in clock cycles using the \emph{\texttt{RDTSCP}} instruction~\cite{RDTSCP}. It does so for each \emph{\texttt{cudaMemcpyPeer()}} execution (initiated by the CPU) that transfers 256 bytes (a single packet) via NVLink. (2) Program \textbf{B}, which transfers varying sizes of data over the shared NVLink, either explicitly via \emph{\texttt{cudaMemcpyPeer()}} or implicitly through Unified Virtual Memory.
We conduct experiments on both DGX-1 (NVLink-V1) and GCP machines (NVLink-V2).

\begin{figure}[tbh]
    \centering
    \includegraphics[width=0.45\textwidth]{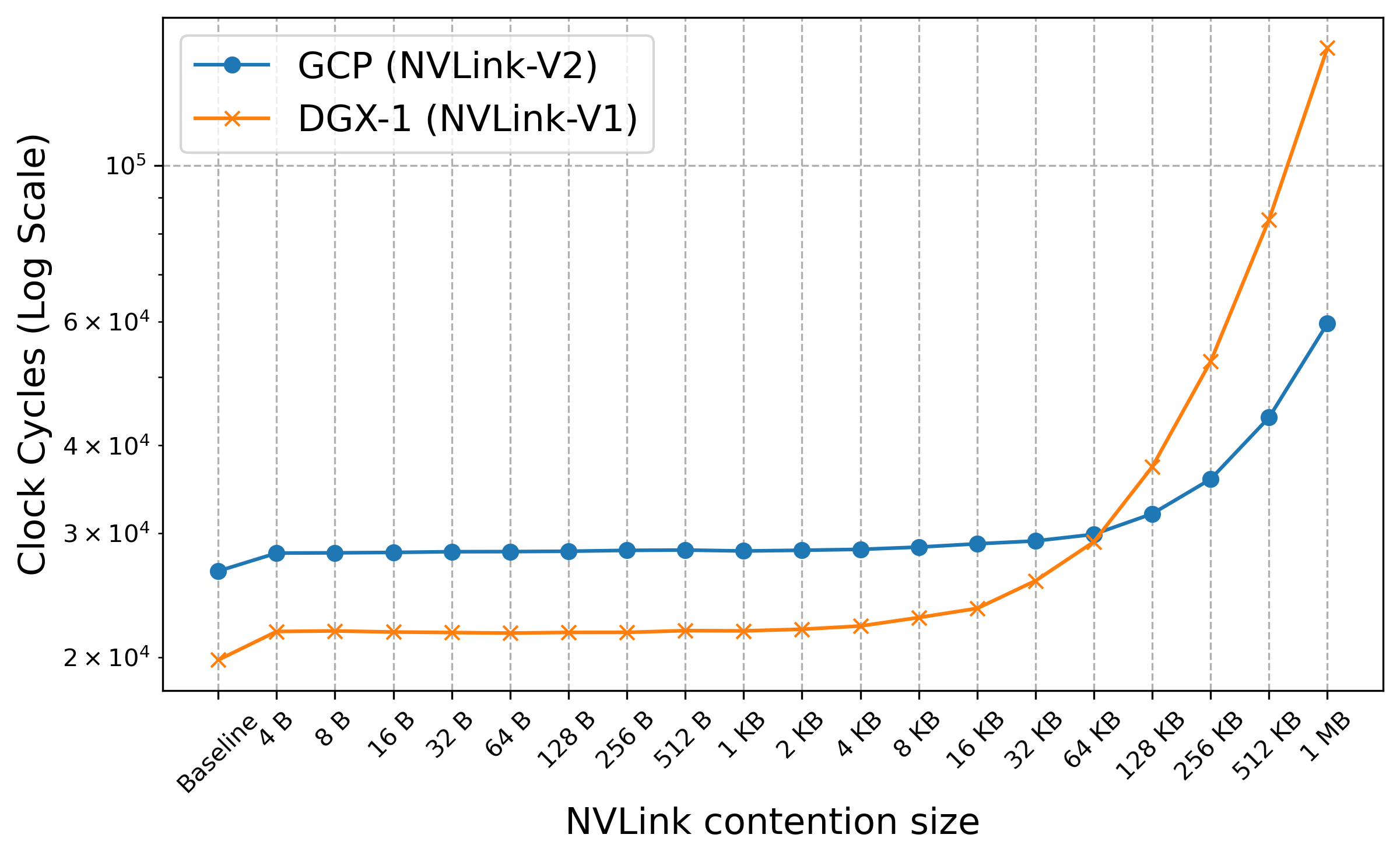}
    \caption{NVLink contention measurements.}
    \label{fig:nvlink_contention_measurement}
\end{figure}



\noindent \textbf{{Contention size effect.}} Fig.{~\ref{fig:nvlink_contention_measurement}} illustrates the impact of contention size on both NVLink-V1 and NVLink-V2 when program \textbf{B} transfers data through NVLink explicitly using \texttt{cudaMemcpyPeer()}. 
When two programs share the NVLink, even a data transfer as small as 4 bytes can introduce a measurable delay exceeding 10\%. As the contention size increases, we observe a corresponding increase in timing delay. When the contention data size exceeds 256 bytes, both NVLink-V1 and NVLink-V2 experience consistent and noticeable contention delays. 

We also observe that the interconnect topology in multi-GPU systems significantly impacts contention. As illustrated in Fig.{~\ref{fig:threat_model}}, a DGX system (NVLink-v1) connects two P100 GPUs via a single NVLink slot, while Google Cloud Platform (NVLink-v2) connects two V100 GPUs via three NVLink slots. Our measurements (see Fig.{~\ref{fig:nvlink_contention_measurement}}) show that single-slot configurations are more prone to contention than multi-slot setups.



\begin{tcolorbox}[colback=gray!10,colframe=black,boxrule=0.5pt,arc=2mm,outer arc=2mm,]
\textit{\textbf{Observation 5}:  
NVLink contention leads to observable increases in data transfer time.}
\end{tcolorbox}

\noindent \textbf{Summary of identified NVLink leakages.} Observations 2, 3, and 5 reveal that NVLink performance counters, although not intended for this purpose, aggregate data transaction patterns from all users on the shared NVLink, which can be exploited. These patterns include transfer sizes, directions, and timings, which an attacker can use to infer a victim’s behavior. Specifically:
\begin{itemize}
    \item Observation 2: By examining NVLink receive/transmit values, an attacker can determine the victim’s data transfer direction.
    \item Observation 3: By tracking NVLink total counters, an attacker can estimate the size of the victim’s transactions.
    \item Observation 5: By introducing NVLink contention, an attacker can exploit a timing-based side channel to gather further information about the victim’s activities.
\end{itemize}




\section{Intra-VM Covert-Channel Attacks}
\label{sec:covert_channel}

We explore two designs of covert channels: {\em ContenLink} and  {\em LeakyCounterLink}.  {\em ContenLink} is based on timing variations due to contention effects on the shared links. {\em LeakyCounterLink} exploits NVLink's leaky performance counters.



\subsection{Covert-Channel Design}
This section describes the design of the two covert channels.  Important to covert channels is the ability of the sender and receiver to synchronize, which we describe first.  

\begin{figure}[tbh]
    \centering
    \includegraphics[width=0.40\textwidth]{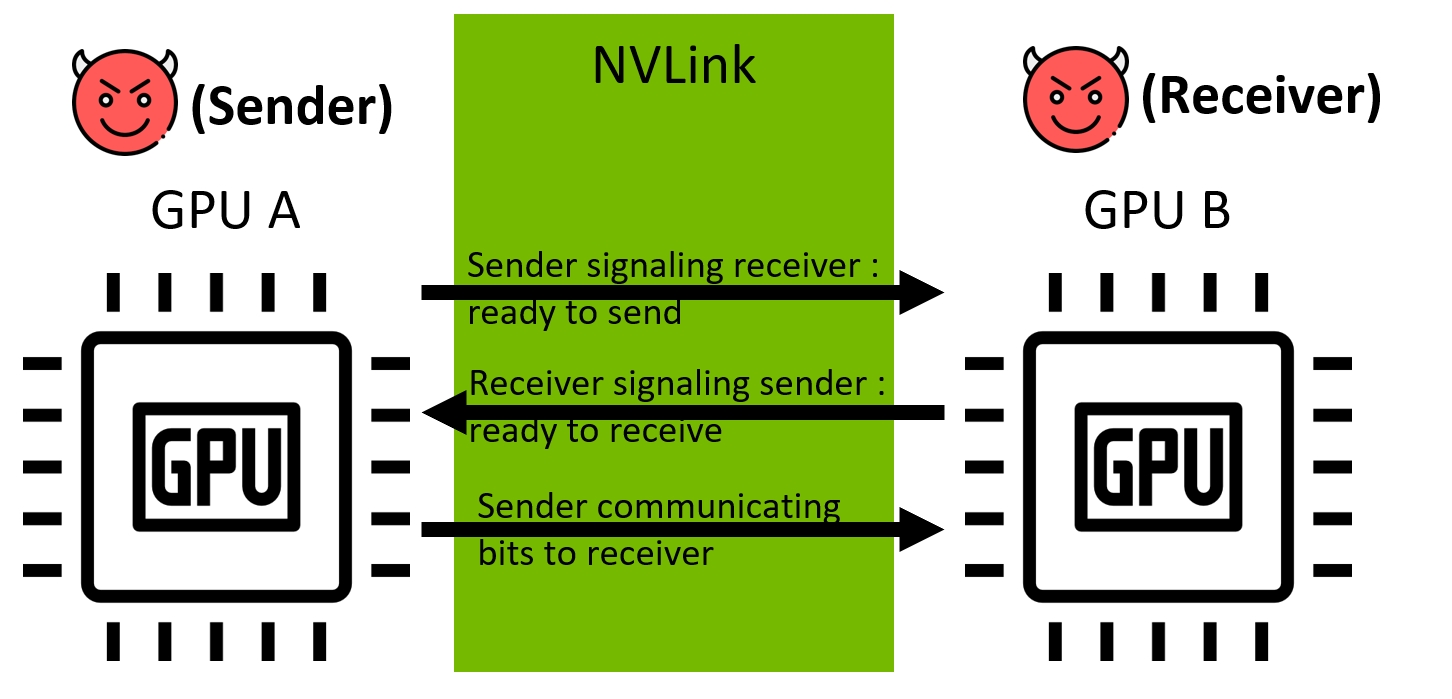}
    \caption{Synchronization protocol for covert channels.}
    \label{fig:covert_channel}
\end{figure}

\noindent \textbf{{Synchronization.}} Synchronization is essential for improving the bandwidth and controlling the error rate on a covert channel.   Fig.~\ref{fig:covert_channel} illustrates the synchronization process which is conducted in three phases.  The sender begins the handshake by transmitting a pre-agreed data pattern through NVLink, thereby signaling its readiness to send. 
Upon detecting this signal via GPU performance counters, the receiver program acknowledges its readiness to receive by reciprocating with a data transfer of equivalent size back to the sender’s GPU. This completes the second step of the handshake. Once the handshake is successfully concluded, the sender proceeds to transmit the covert data bits to the spy program. 

\noindent \textbf{{Design 1: ContenLink}.} As we observed in Section~\ref{sec:nvlink_analysis}, contention on a shared NVLink leads to an increase in data transfer time. We build a covert channel that exploits this timing difference to transmit data covertly.  Algorithm~\ref{alg:conten_receiver} outlines the design of the \emph{ContenLink} receiver. Initially, the receiver allocates two 256-byte arrays, $R_0$ in the local GPU 0 and $R_1$ in the remote GPU 1, respectively. 
We set the array size to 256 bytes because, based on our first observation, this is the maximum size of a single NVLink packet.  
The receiver measures the execution time for invoking the \texttt{cudaMemcpyPeer()} API, which facilitates the copying of arrays $R_1$ to $R_0$ via NVLink, using the hardware timer accessible through the \texttt{RDTSCP} instruction. If the measured time falls below a threshold $T$, the receiver interprets this as free of contention on the NVLink, indicating a received bit of `0'. Conversely, a measurement exceeding the threshold signifies contention, interpreted as a received bit of `1'.

\begin{algorithm}[tbh]
\caption{Receiver for ContenLink Covert Channel} \label{alg:conten_receiver}
\begin{small}

// $D_{\text{receive}}[N]$ is an array of $N$ bits to receive a message\;
// T is the threshold to distinguish between bits `1' and `0'\;
     Allocate an array $R_0$ in local GPU 0's memory\;
     Allocate an array $R_1$ in remote GPU 1's memory\;
     Synchronization()\;
        
    \For{$i \gets 0$ \KwTo $N-1$}{
        Record time \textit{start} via \texttt{RDTSCP}\;
        Execute \texttt{cudaMemcpyPeer()} to copy $R_1$ to $R_0$\;
        Record time \textit{end} via \texttt{RDTSCP}\;
        
        \eIf{\(end - start < T\)}
        {
        \(D_{\text{receive}}[i] \gets 0\);
        }
        {
        \(D_{\text{receive}}[i]\gets 1\);
        }       
            
    }
\end{small}
\end{algorithm}

Algorithm~\ref{alg:sender} presents the sender's design for the covert channel.  The initial step involves allocating two arrays, $S_0$ and $S_1$, on two GPUs. The differing sizes of these arrays are intended to exert varying levels of contention pressure on the NVLink.
We evaluate the impact of varying sender sizes on the covert channel's bandwidth and error rate in Section~\ref{subsec:covert_evaluation}. 
After synchronizing with the receiver, the sender starts the transmission of covert messages.  To transfer a bit `1', it uses \emph{\texttt{cudaMemcpyPeer()}} API to copy $S_1$ to $S_0$ from remote GPU to local GPU, creating contention on the shared NVLink with the receiver. For transmitting a bit `0', it performs $K$ loops of \emph{\texttt{NOP}} instructions, which does not cause contention on the NVLink.


\begin{algorithm}[tbh]
\caption{Sender for Covert Channel} \label{alg:sender}
\begin{small}

// $D_{\text{sender}}[N]$ is an array of $N$ bits used to send a message\;
// $K$ is the number of execution of NOPs\;
     Allocate an array $S_0$ in local GPU 0's memory\;
     Allocate an array $S_1$ in remote GPU 1's memory\;
     Synchronization()\;
        
    \For{$i \gets 0$ \KwTo $N-1$}{

        \eIf{\(D_{\text{receive}}[i] == 1\)}
        {
        Execute \texttt{cudaMemcpyPeer()} to copy $S_1$ to $S_0$\;
        }
        {
        \For{$j \gets 0$ \KwTo $K - 1$}{
            Execute a \texttt{NOP};
            }
        }   
        

    }
\end{small}
\end{algorithm}

\noindent \textbf{{Design 2: LeakyCounterLink}.} Based on our second and third observations, we develop an alternative design for an intra-VM covert channel. 
By monitoring two NVLink counters: \texttt{nvlink\_total\_da\-ta\_transmitted} and \texttt{nvlink\_total\_data\_received}, the receiver can observe not only its own data transfer patterns but also those of the sender. 
Algorithm~\ref{alg:leakyctr_receiver} shows the design of the receiver. Similar to \emph{ContenLink} receiver, it also allocates two 256-byte arrays: one on local GPU 0 and the other on remote GPU 1.
After synchronizing with the sender, the receiver continuously profiles two NVLink counters while executing the \emph{\texttt{cudaMemcpyPeer()}} API. If the counter values are less than the threshold $T$, it records a bit of `0'. Otherwise, a bit of `1' is stored in the receiver's array.



\begin{algorithm}[tbh]
\caption{Receiver for LeakyCounterLink Covert Channel} \label{alg:leakyctr_receiver}
\begin{small}

// $D_{\text{receive}}[N]$ is an array of $N$ bits used to receive a message\;
// T is the threshold to differentiate bits `1' and `0'\;
     Allocate an array $R_0$ in local GPU 0's memory\;
     Allocate an array $R_1$ in remote GPU 1's memory\;
     Synchronization()\;
        
    \For{$i \gets 0$ \KwTo $N-1$}{
        \texttt{cudaProfilerStart()}\;
        Execute \texttt{cudaMemcpyPeer()} to copy $R_1$ to $R_0$\;
        \texttt{cudaProfilerStop()}\;
        Record NVLink leaky counters value to \textit{counter}\;      
        \eIf{\(counter < T\)}
        {
        \(D_{\text{receive}}[i] \gets 0\);
        }
        {
        \(D_{\text{receive}}[i]\gets 1\);
        }       
            
    }
\end{small}
\end{algorithm}

\subsection{Covert Channel Evaluation}
\label{subsec:covert_evaluation}

In this section, we evaluate two designs of intra-VM covert-channel attacks on both DGX-1 system and GCP machines. 

\begin{figure}[tbh]
    \centering
    \includegraphics[width=0.45\textwidth]{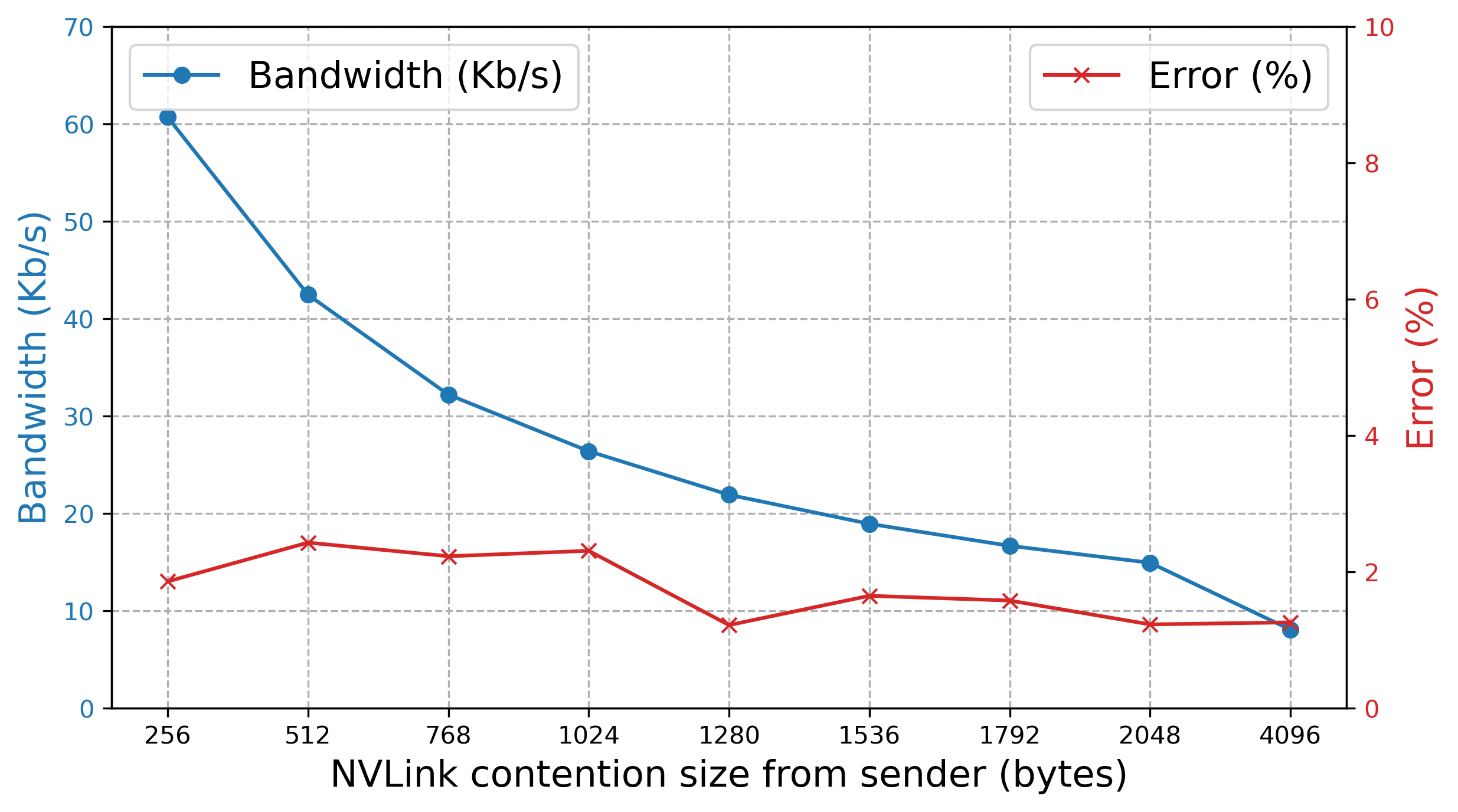}
    \caption{DGX contention covert channel results.}
    \label{fig:dgx_contention_result}
\end{figure}

\begin{figure}[tbh]
    \centering
    \includegraphics[width=0.45\textwidth]{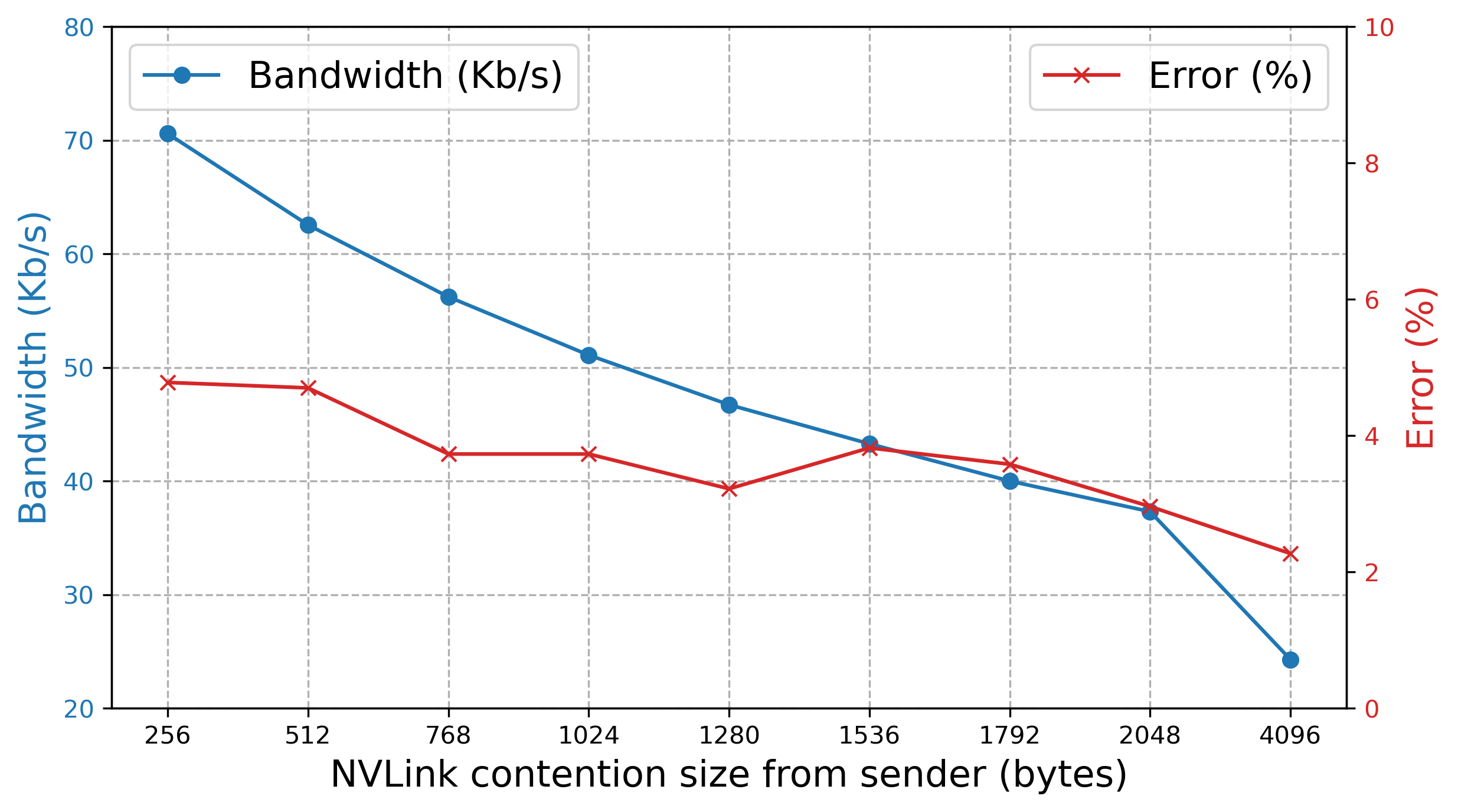}
    \caption{GCP contention covert channel results.}
    \label{fig:gcp_contention_result}
\end{figure}

\noindent \textbf{{Evaluation of \textit{ContenLink}.}} To evaluate \textit{ContenLink}, we transmitted a 10,000-bit message five times and calculated the average bandwidth and error rate. The error rate was determined using the Levenshtein edit distance~\cite{miller2009levenshtein}. Each message, composed of an equal number of `0' and `1' bits, was randomly generated.
Fig.~\ref{fig:dgx_contention_result} and Fig.~\ref{fig:gcp_contention_result} illustrate the bandwidth and error rate on two platforms. Based on our contention measurements detailed in Section~\ref{subsec:contention_nvlink}, the varying data pressures exerted by the sender distinctly affect the covert channel's performance.
We tested sender data sizes ranging from 256 bytes to 4 MB. We ignore sizes smaller than 256 bytes because we find that when the contention size is small, it becomes challenging for the receiver to distinguish between bit `1' and `0', resulting in a high error rate.
On the GCP machine, we achieved the highest bandwidth of 70.59 Kb/s with a sender data size of 256 bytes, accompanied by an error rate of 4.78\%. Similarly, on the DGX-1 machine, the highest bandwidth reached was 60.71 Kb/s with the same sender data size, resulting in a lower error rate of 1.86\%.
As the data pressure from the sender increases, we observe a corresponding decrease in the bandwidth of covert channels on both GCP and DGX systems, although the error rate remains stable.

\noindent \textbf{{Evaluation of \textit{LeakyCounterLink}.}} We employed the same evaluation methods for \textit{LeakyCounterLink} as were used for \textit{ContenLink}. However, its performance was inferior to that of \textit{ContenLink}. On the GCP platform, the highest bandwidth achieved was 1.88 Kb/s with a sender data size of 256 bytes, accompanied by an error rate of 7.50\%. Similarly, on the DGX-1 machine, the highest bandwidth recorded was 1.39 Kb/s with the same sender data size, resulting in an error rate of 8.80\%. 
When we use CUPTI to profile leaky counters, the CUPTI profiler introduces significant overhead. This is due to the need for synchronization of profiling resources between the host and the device, as well as the delivery of activity buffers to the client via the buffer completed callback~\cite{CUPTI}. Consequently, this overhead adversely affects the bandwidth of \textit{LeakyCounterLink}.
Additionally, although the leaky counters can profile the transaction behavior of other users, they are unstable and noisy. This instability contributes to a higher error rate compared to \textit{ContenLink}.

\section{Intra-VM Side-Channel Attacks}
\label{sec:side_channel}
In this section, we demonstrate two end-to-end side-channel attacks: application fingerprinting and 3D object fingerprinting attacks.

\subsection{Attack 1: Application Fingerprinting}
\label{subsec:app_finger}

In this attack, we reveal that an adversary can infer the specific HPC applications or deep learning models by exploiting NVLink side-channel leakages. We test application fingerprint attacks on 8 HPC applications from the OpenMM~\cite{eastman2017openmm} benchmarks and 10 deep-learning models as the victim applications. 


\noindent \textbf{{OpenMM benchmark.}} OpenMM is a high-performance toolkit tailored for molecular dynamics simulations, supporting multi-GPU systems. In this study, we center our attention on 8 benchmark applications~\cite{openmmbenchmark}: rf, pme, apoa1-rf, apoa1-pme, apoa1-ljpme, amoeba-pme, amber20-dhfr, and amber20-cellulose. 

\noindent \textbf{{Deep learning models.}} 
In this work, we consider the attacker's attempts to extract the model structure like in previous work~\cite{naghibijouybari2018rendered, wei2020leaky, tan2021invisible, dutta2023spy}. 
Specifically, we select 10 popular deep learning models (5 simple and 5 complex). 
The simple models include a Multi-layer Perceptron (MLP), two basic Convolutional Neural Networks (CNNs), a regression model, and a Long Short-Term Memory (LSTM) network~\cite{hochreiter1997long}.  For the complex models, we chose 5 famous models: AlexNet~\cite{krizhevsky2012imagenet}, VGG16~\cite{simonyan2014very}, GoogLeNet~\cite{szegedy2015going}, and two variants of ResNet (ResNet-18 and ResNet-50)~\cite{he2016deep} used in Computer Vision (CV).
Each model was trained on the MNIST dataset~\cite{deng2012mnist}, utilizing PyTorch-supported data parallelism for distributed training across 100 iterations~\cite{pytorch}. The batch size was set at 64. For complex models, the image size was resized to 224 by 224.
We leave the layer hyper-parameter extraction as future work.

\noindent \textbf{{Experimental setup and data collection.}} 
We assume the victim deploys her application using three different GPU distribution strategies, with 2, 4, or 8 GPU configurations. 
The spy application continuously monitors the side-channel leakages from a single NVLink utilized by the victim.
The background spy program was set up to profile the counter values: \texttt{nvlink\_total\_data\_received}, along with timing delays caused by the victim application’s contention on NVLink. 
As outlined in Section~\ref{subsec:available_leakage_vector}, we turn off the aggregation mode of the profiler and collect leakage for each individual link. 
Consequently, for the DGX-1 platform, which encompasses 4 NVLink-V1 slots per GPU, the spy program collects 5 leakage vectors, encompassing 4 vectors for links and timing delay information. 
Similarly, with the GCP platform, which features 6 NVLink-V2 slots per GPU, attackers can acquire 7 leakage vectors. This includes 6 vectors from links and timing delay. During the data collection phase, we collected 50 traces of side-channel leakages for each application. The dataset was divided into training and testing sets using an 80/20 split ratio.

\begin{figure}[tbh]
    \centering
    \begin{subfigure}[b]{0.42\textwidth}
        \centering
        \includegraphics[width=\textwidth]{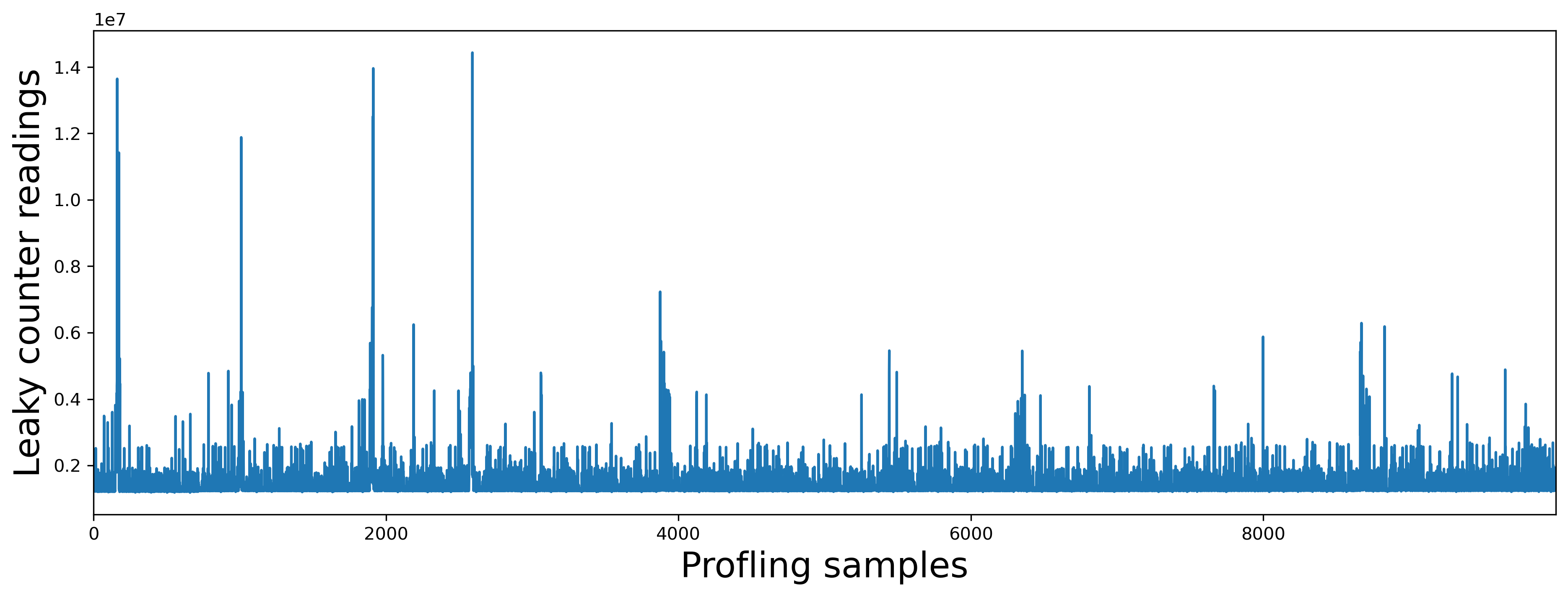}
        \caption{The NVLink leakage traces of rf.}
        \label{fig:rf}
    \end{subfigure}
    \hfill 
    \begin{subfigure}[b]{0.42\textwidth}
        \centering
        \includegraphics[width=\textwidth]{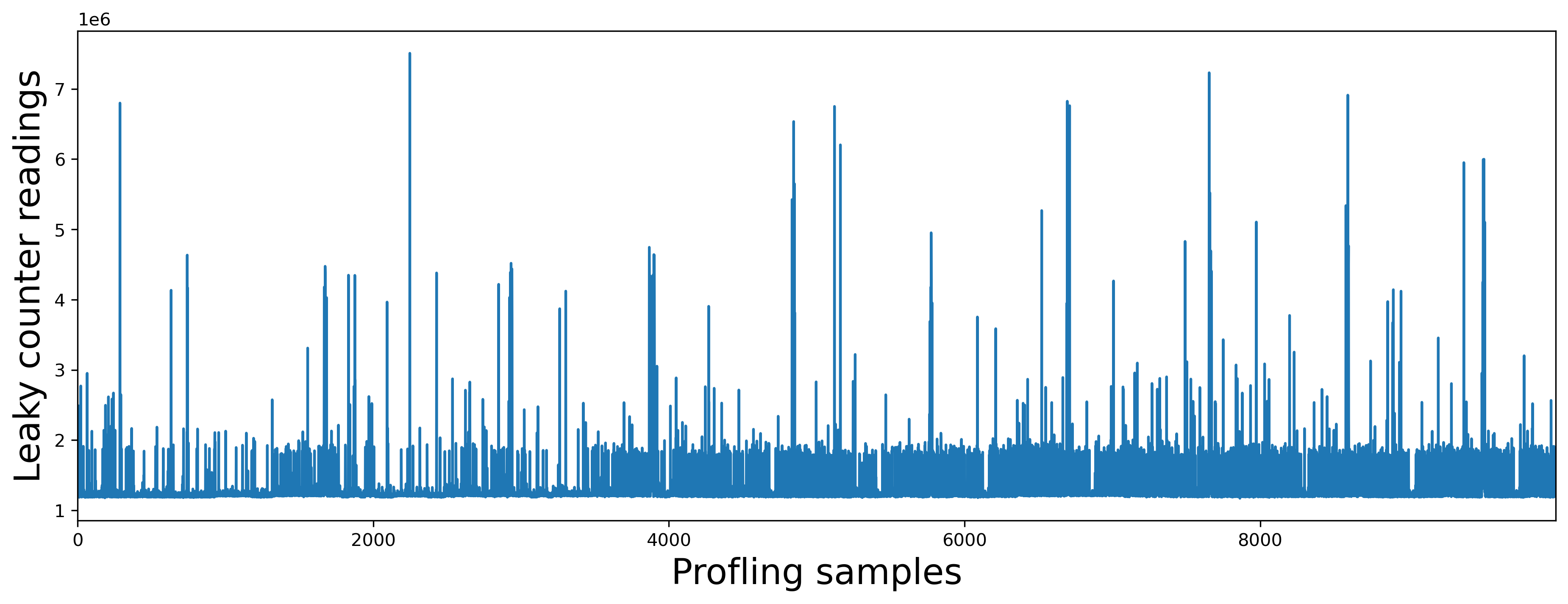}
        \caption{The NVLink leakage traces of amber20\_cellulose.}
        \vspace{10pt}
        \label{fig:amber20}
    \end{subfigure}
    \hfill
    \begin{subfigure}[b]{0.42\textwidth}
        \centering
        \includegraphics[width=\textwidth]{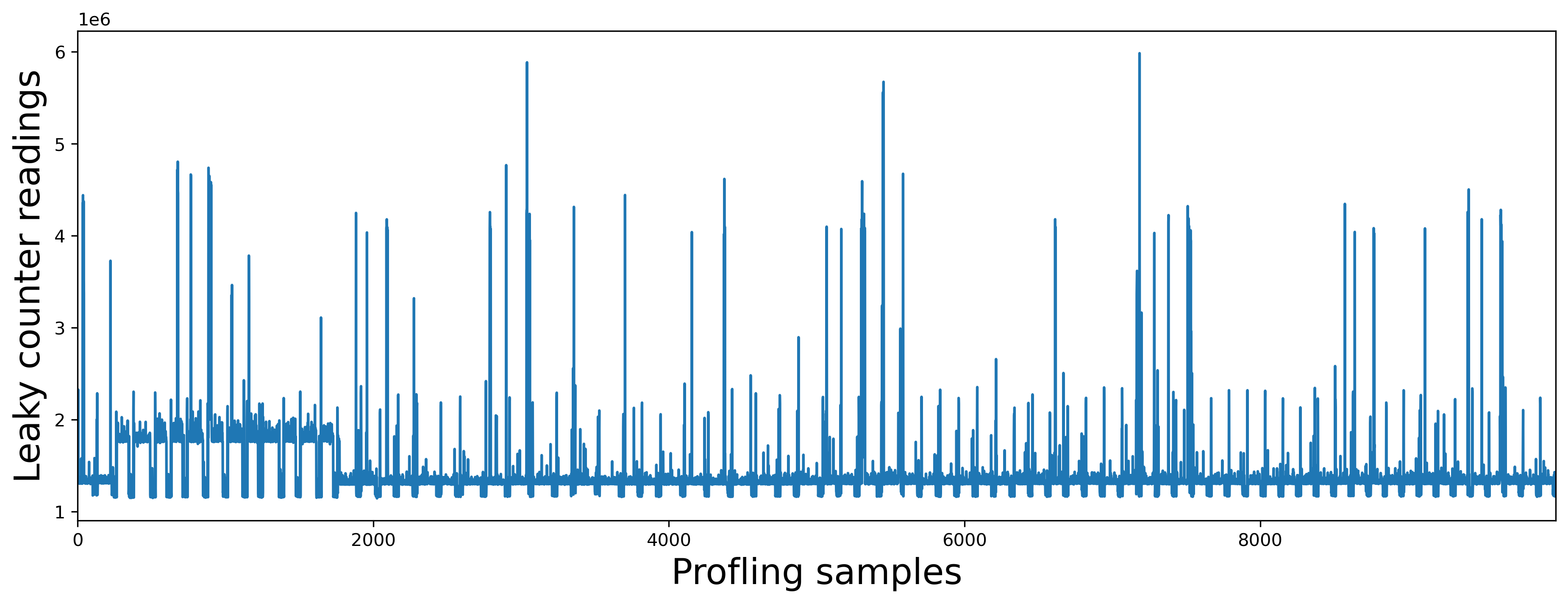}
        \caption{The NVLink leakage traces of AlexNet.}
        \label{fig:alexnet}
    \end{subfigure}
    \hfill 
    \begin{subfigure}[b]{0.42\textwidth}
        \centering
        \includegraphics[width=\textwidth]{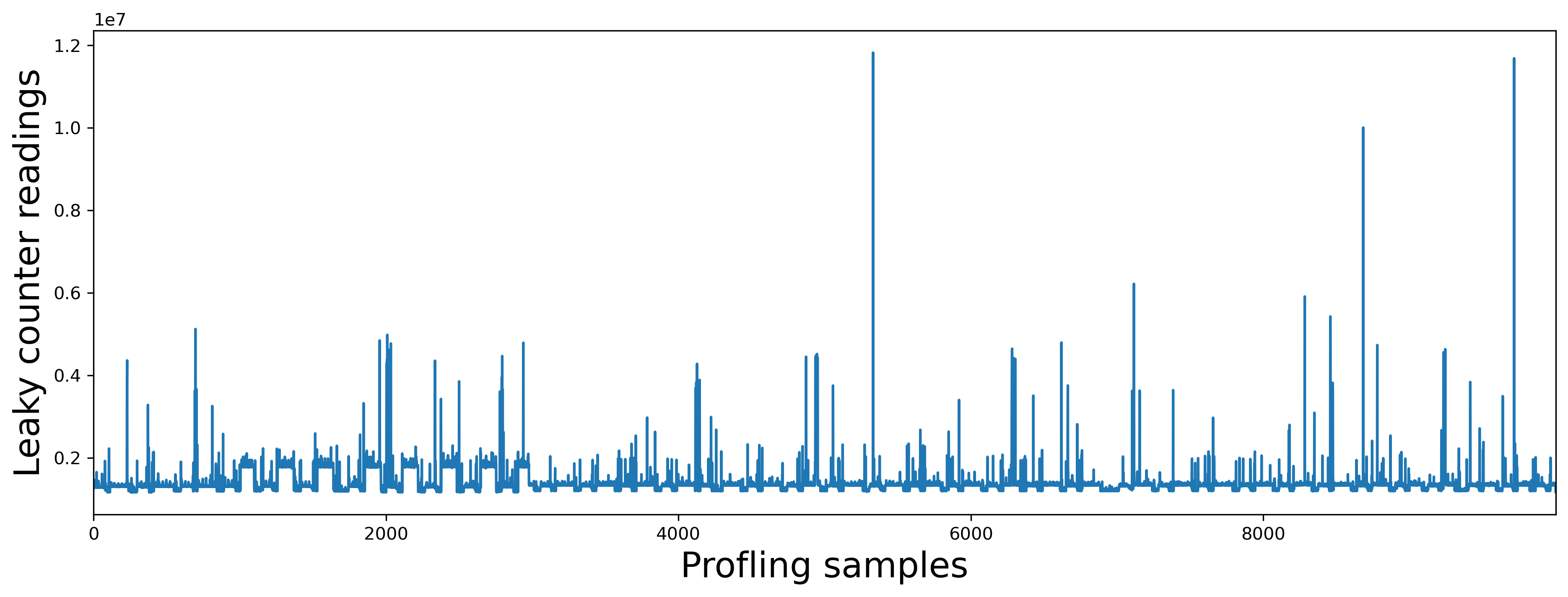}
        \caption{The NVLink leakage traces of ResNet-50.}
        \label{fig:resnet50}
    \end{subfigure}    
    
    \caption{Comparison of NVLink leakage traces.}           
    \label{fig:app_traces}
\end{figure}

\noindent \textbf{{Observing benchmark distinguishability.}} Fig.~\ref{fig:app_traces} displays traces for 4 applications—rf, amber20\_cellulose, AlexNet and ResNet-50—highlighting their distinguishable characteristics. The numbers on the X-axis correspond to the profiling samples. The Y-axis displays the leaky counter readings; in this case, we use \texttt{nvlink\_total\_ data\_received} in bytes.

\noindent \textbf{{Feature engineering and classification.}} 
We extract features from the collected time-series data.
We use a sliding window approach with a window size of 100 to compute 12 statistical features listed in Table~\ref{tb:features_definition}.
We then use the extracted features to build classification models, employing three standard machine learning algorithms: K Nearest Neighbors (KNN)~\cite{knn} with parameter settings: $n\_neighbors = 5, leaf\_size = 30$, XGBoost~\cite{chen2016xgboost} with parameter settings: $n\_estimators = 100, ma\-x\_depth=6$, and Light Gradient Boosting Machine (LightGBM)~\cite{ke2017lightgbm} with parameter settings: $num\_leaves = 31,max\_depth = -1,n\_esti\-mators = 100$. 
To assess the performance of these classifiers, we calculated three metrics \cite{hossin2015review}, including the F1 score (F1), Precision (Prec), and Recall (Rec).

\begin{table}[tbh]
\small
\centering
\caption{Definitions of statistical features used in \textit{NVBleed}. }
\label{tb:features_definition}
\begin{tabular}{|L{1.5cm}|L{6cm}|}
\hline
\textbf{Features} & \textbf{Description}                     \\ \hline \hline
mean                  &   Mean of all values.
\\ \hline
max           &  Maximum of all values.
\\ \hline
min              &  Minimum of all values.
\\ \hline
median             &    Median of all values.
\\ \hline
std            &  Standard deviation. 
\\ \hline
var            & Variance. 
\\ \hline
range             & Difference of maximum and minimum.
\\ \hline
sum             & Sum of all values.
\\ \hline
count\_am            &  Count of observations that exceed the mean.
\\ \hline
percent\_25            & 25th percentile value.
\\ \hline
percent\_75            & 75th percentile value.
\\ \hline
iqr\_val & Interquartile range.
\\ \hline

\end{tabular}

\end{table}

\noindent \textbf{Results.} We evaluated the application fingerprinting attack under two attacker capabilities: (1) \textbf{Timing only}: exploiting contention-based timing leakage on NVLink, and (2) \textbf{Timing + Counters}: combining timing leakage with leaky performance counters.  This second attack is only possible if the performance counters are enabled in the GPU driver.   For 2-GPU configurations, Table{~\ref{tb:app_fingerprint_2gpu}} presents the classification performance results. XGBoost and LightGBM achieve the highest accuracy on the DGX testbed and GCP machine, with F1 scores exceeding 82\% and 96\%, respectively, using only timing leakage. Incorporating leaky performance counters alongside timing leakage further improves classification performance, achieving F1 scores above 92\% and 97\% for the DGX testbed and GCP machine, respectively.
For 4-GPU configurations (Table{~\ref{tb:app_fingerprint_4gpu}}), under timing leakage alone, LightGBM delivers the best performance on DGX (F1 = 88.04\%), while XGBoost narrowly leads on GCP (F1 = 89.17\%). Once performance counters are combined with timing leakage, XGBoost becomes the top performer on both platforms, surpassing 95\% F1 on DGX and 88\% on GCP.
In 8-GPU configurations (Table{~\ref{tb:app_fingerprint_8gpu}}), with timing leakage only, LightGBM attains the highest F1 score on both DGX (82.27\%) and GCP (73.41\%). Incorporating leaky performance counters further increases LightGBM’s scores to 90.91\% on DGX and 85.40\% on GCP.

As the multi-GPU system topology becomes more complex with additional GPUs, the overall accuracy of the classifiers decreases. However, combining contention-based timing signals with performance counter data significantly enhances classification accuracy while maintaining strong attack performance, with F1 scores consistently exceeding 85\%.


\begin{table}[tbh]
\footnotesize
\centering
\caption{Application fingerprint performance: F1 (\%), Precision (\%), and Recall (\%) on DGX and GCP (2 GPUs). \timeicon = Contention-based timing leakage, \faCalculator = Leaky performance counters.}

\begin{tabular}{cc|ccc||ccc|}
\cline{3-8}
 &  & \multicolumn{3}{c||}{\textbf{DGX}} & \multicolumn{3}{c|}{\textbf{GCP}} \\ 
\cline{3-8} 
\textbf{} &
  \textbf{} &
  \multicolumn{1}{c|}{\textbf{F1}} &
  \multicolumn{1}{c|}{\textbf{Prec}} &
  \textbf{Rec} &
  \multicolumn{1}{c|}{\textbf{F1}} &
  \multicolumn{1}{c|}{\textbf{Prec}} &
  \textbf{Rec} \\ 
\hline \hline
\multicolumn{1}{|c|}{\multirow{3}{*}{\textbf{\timeicon}}} &
  KNN &
  \multicolumn{1}{c|}{36.22} &
  \multicolumn{1}{c|}{37.26} &
  39.44 &
  \multicolumn{1}{c|}{55.77} &
  \multicolumn{1}{c|}{55.97} &
  58.89 \\ 
\cline{2-8} 
\multicolumn{1}{|c|}{} &
  XGBoost &
  \multicolumn{1}{c|}{\textbf{82.69}} &
  \multicolumn{1}{c|}{82.95} &
  82.78 &
  \multicolumn{1}{c|}{96.65} &
  \multicolumn{1}{c|}{96.90} &
  96.67 \\ 
\cline{2-8} 
\multicolumn{1}{|c|}{} &
  LightGBM &
  \multicolumn{1}{c|}{81.21} &
  \multicolumn{1}{c|}{81.25} &
  81.67 &
  \multicolumn{1}{c|}{\textbf{96.77}} &
  \multicolumn{1}{c|}{97.23} &
  96.67 \\ 
\hline \hline
\multicolumn{1}{|c|}{\multirow{3}{*}{\textbf{\timecounter}}} &
  KNN &
  \multicolumn{1}{c|}{25.96} &
  \multicolumn{1}{c|}{31.45} &
  26.11 &
  \multicolumn{1}{c|}{55.77} &
  \multicolumn{1}{c|}{55.97} &
  58.89 \\ 
\cline{2-8} 
\multicolumn{1}{|c|}{} &
  XGBoost &
  \multicolumn{1}{c|}{90.87} &
  \multicolumn{1}{c|}{91.45} &
  91.11 &
  \multicolumn{1}{c|}{\textbf{97.78}} &
  \multicolumn{1}{c|}{98.06} &
  97.78 \\ 
\cline{2-8} 
\multicolumn{1}{|c|}{} &
  LightGBM &
  \multicolumn{1}{c|}{\textbf{92.22}} &
  \multicolumn{1}{c|}{93.12} &
  92.22 &
  \multicolumn{1}{c|}{96.10} &
  \multicolumn{1}{c|}{96.93} &
  96.11 \\ 
\hline
\end{tabular}
\label{tb:app_fingerprint_2gpu}
\end{table}

\begin{table}[tbh]
\footnotesize
\centering
\caption{Application fingerprint performance: F1 (\%), Precision (\%), and Recall (\%) on DGX and GCP (4 GPUs). \timeicon = Contention-based timing leakage, \faCalculator = Leaky performance counters.}

\begin{tabular}{cc|ccc||ccc|}
\cline{3-8}
 &
   &
  \multicolumn{3}{c||}{\textbf{DGX}} &
  \multicolumn{3}{c|}{\textbf{GCP}} \\ \cline{3-8} 
\textbf{} &
  \textbf{} &
  \multicolumn{1}{c|}{\textbf{F1}} &
  \multicolumn{1}{c|}{\textbf{Prec}} &
  \textbf{Rec} &
  \multicolumn{1}{c|}{\textbf{F1}} &
  \multicolumn{1}{c|}{\textbf{Prec}} &
  \textbf{Rec} \\ \hline \hline
\multicolumn{1}{|c|}{\multirow{3}{*}{\textbf{\timeicon}}} &
  KNN &
  \multicolumn{1}{c|}{54.20} &
  \multicolumn{1}{c|}{58.26} &
  57.22 &
  \multicolumn{1}{c|}{66.65} &
  \multicolumn{1}{c|}{69.16} &
  67.78 \\ \cline{2-8} 
\multicolumn{1}{|c|}{} &
  XGBoost &
  \multicolumn{1}{c|}{85.77} &
  \multicolumn{1}{c|}{86.51} &
  85.56 &
  \multicolumn{1}{c|}{\textbf{89.17}} &
  \multicolumn{1}{c|}{89.56} &
  89.44 \\ \cline{2-8} 
\multicolumn{1}{|c|}{} &
  LightGBM &
  \multicolumn{1}{c|}{\textbf{88.04}} &
  \multicolumn{1}{c|}{88.64} &
  87.78 &
  \multicolumn{1}{c|}{89.05} &
  \multicolumn{1}{c|}{89.36} &
  89.44 \\ \hline \hline
\multicolumn{1}{|c|}{\multirow{3}{*}{\textbf{\timecounter}}} &
  KNN &
  \multicolumn{1}{c|}{21.67} &
  \multicolumn{1}{c|}{23.73} &
  24.44 &
  \multicolumn{1}{c|}{63.26} &
  \multicolumn{1}{c|}{66.50} &
  64.44 \\ \cline{2-8} 
\multicolumn{1}{|c|}{} &
  XGBoost &
  \multicolumn{1}{c|}{\textbf{95.43}} &
  \multicolumn{1}{c|}{95.70} &
  95.56 &
  \multicolumn{1}{c|}{\textbf{88.60}} &
  \multicolumn{1}{c|}{89.79} &
  88.60 \\ \cline{2-8} 
\multicolumn{1}{|c|}{} &
  LightGBM &
  \multicolumn{1}{c|}{92.64} &
  \multicolumn{1}{c|}{93.20} &
  92.78 &
  \multicolumn{1}{c|}{88.14} &
  \multicolumn{1}{c|}{89.08} &
  88.33 \\ \hline
\end{tabular}
\label{tb:app_fingerprint_4gpu}
\end{table}

\begin{table}[tbh]
\footnotesize
\centering
\caption{Application fingerprint performance: F1 (\%), Precision (\%), and Recall (\%) on DGX and GCP (8 GPUs). \timeicon = Contention-based timing leakage, \faCalculator = Leaky performance counters.}

\begin{tabular}{cc|ccc||ccc|}
\cline{3-8}
 &
   &
  \multicolumn{3}{c||}{\textbf{DGX}} &
  \multicolumn{3}{c|}{\textbf{GCP}} \\ \cline{3-8} 
\textbf{} &
  \textbf{} &
  \multicolumn{1}{c|}{\textbf{F1}} &
  \multicolumn{1}{c|}{\textbf{Prec}} &
  \textbf{Rec} &
  \multicolumn{1}{c|}{\textbf{F1}} &
  \multicolumn{1}{c|}{\textbf{Prec}} &
  \textbf{Rec} \\ \hline \hline
\multicolumn{1}{|c|}{\multirow{3}{*}{\textbf{\timeicon}}} &
  KNN &
  \multicolumn{1}{c|}{42.38} &
  \multicolumn{1}{c|}{47.09} &
  44.44 &
  \multicolumn{1}{c|}{63.01} &
  \multicolumn{1}{c|}{66.60} &
  63.33 \\ \cline{2-8} 
\multicolumn{1}{|c|}{} &
  XGBoost &
  \multicolumn{1}{c|}{79.30} &
  \multicolumn{1}{c|}{79.40} &
  79.44 &
  \multicolumn{1}{c|}{71.54} &
  \multicolumn{1}{c|}{71.71} &
  72.78 \\ \cline{2-8} 
\multicolumn{1}{|c|}{} &
  LightGBM &
  \multicolumn{1}{c|}{\textbf{82.27}} &
  \multicolumn{1}{c|}{82.88} &
  82.22 &
  \multicolumn{1}{c|}{\textbf{73.41}} &
  \multicolumn{1}{c|}{74.32} &
  73.89 \\ \hline \hline
\multicolumn{1}{|c|}{\multirow{3}{*}{\textbf{\timecounter}}} &
  KNN &
  \multicolumn{1}{c|}{23.51} &
  \multicolumn{1}{c|}{28.83} &
  25.56 &
  \multicolumn{1}{c|}{66.36} &
  \multicolumn{1}{c|}{69.96} &
  66.67 \\ \cline{2-8} 
\multicolumn{1}{|c|}{} &
  XGBoost &
  \multicolumn{1}{c|}{89.46} &
  \multicolumn{1}{c|}{90.31} &
  89.44 &
  \multicolumn{1}{c|}{84.51} &
  \multicolumn{1}{c|}{85.98} &
  85.00 \\ \cline{2-8} 
\multicolumn{1}{|c|}{} &
  LightGBM &
  \multicolumn{1}{c|}{\textbf{90.91}} &
  \multicolumn{1}{c|}{92.34} &
  91.11 &
  \multicolumn{1}{c|}{\textbf{85.40}} &
  \multicolumn{1}{c|}{86.10} &
  86.11 \\ \hline
\end{tabular}
\label{tb:app_fingerprint_8gpu}
\end{table}

\subsection{Attack 2: Identifying Rendered Characters}
\label{subsec:3d_finger}
We demonstrate a proof-of-concept attack that a spy application can identify which 3D production character the victim is rendering. 
We begin by providing background on how the Blender renders 3D characters. Following this, we outline our attack methodology and discuss the evaluation results.

\noindent \textbf{{3D character rending on multi-GPU.}} 3D scene rendering is an intensive process utilized to generate images and movies from scenes modeled in specialized software environments, such as Blender~\cite{blender}, Unity~\cite{unity}, Unreal Engine~\cite{unrealengine}, and others. 
Utilizing GPUs to accelerate 3D rendering has become a standard practice within these toolkits~\cite{jarovs2021gpu}. The initial step in rendering 3D scenes with GPU assistance involves transferring scene data from CPU to GPU memory.
After the computation is completed, the rendered data is transferred back to the CPU to finish rendering. 
In this study, we choose Blender as the target 3D rendering toolkit for two main reasons: first, its widespread use and open-source nature; second, its support for accelerated rendering on multiple GPUs through NVLink.
By activating the "Distribute Memory Across Devices" option on the Blender, the 3D scene data are not duplicated across each GPU. Instead, this data is transferred to a single GPU from the CPU via PCIe. Subsequently, all NVLink-connected GPUs can share the scene data efficiently. This approach significantly enhances the speed of rendering complex and massive scenes.

\begin{figure}[tbh]
    \centering
    \includegraphics[width=0.5\textwidth]{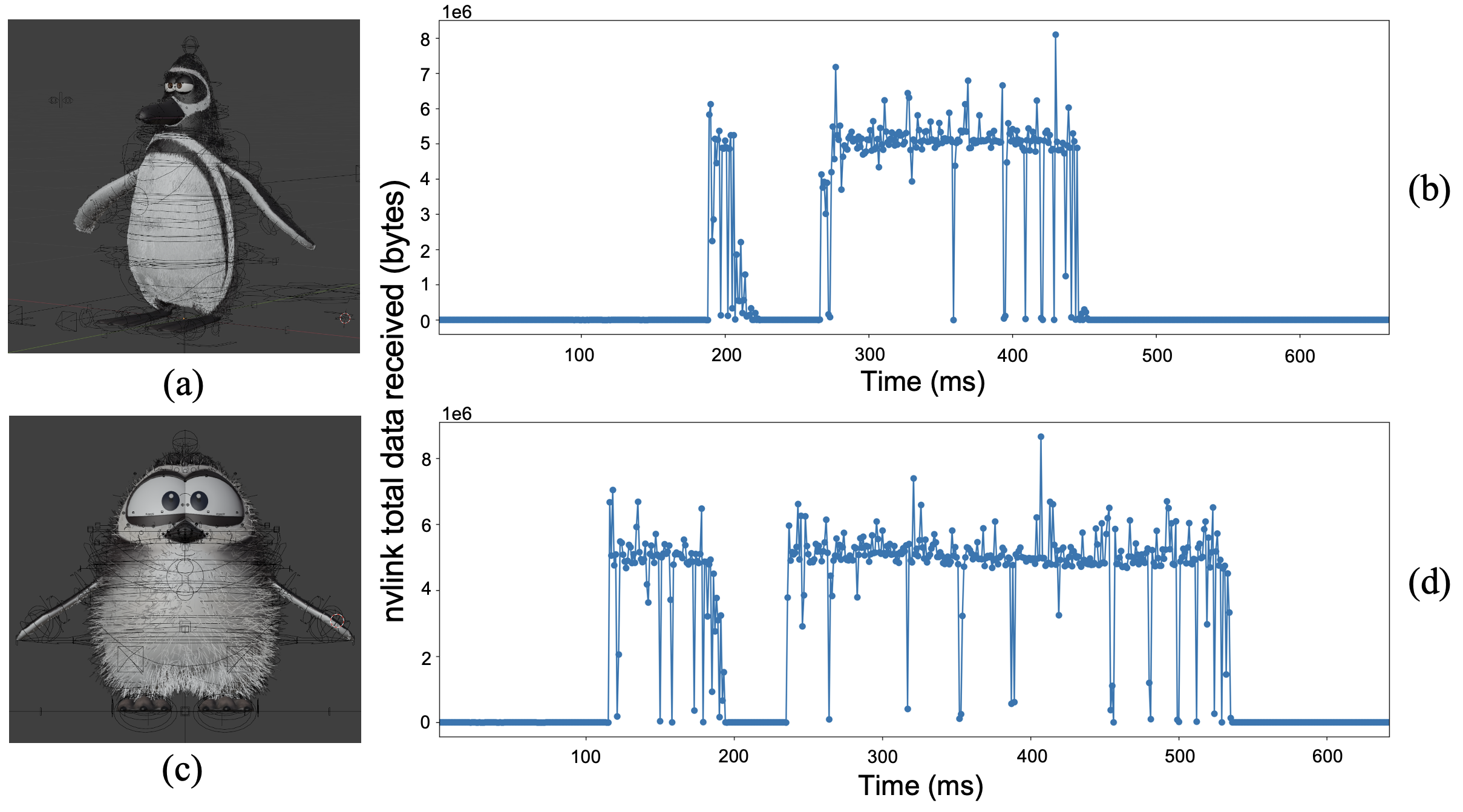}
    \caption{The NVLink leakage traces for two 3D characters from Blender Studio: (a) Character 1: Pinguino~\cite{Pinguino}, (b) the NVLink leakages for Pinguino, (c) Character 2: Oti~\cite{Oti}, and (d) the NVLink leakages for Oti.}
    \label{fig:blender_traces}
\end{figure}

\noindent \textbf{{Observing 3D character distinguishability.}} Fig.~\ref{fig:blender_5_frame} depicts the NVLink counter readings for 5 consecutive frames. These readings clearly show the start and end times of rendering for each frame.
When zoomed in, as illustrated in Fig.~\ref{fig:blender_traces},
we observe that the renderings of two 3D characters (Character 1: Pinguino~\cite{Pinguino} and Character 2: Oti~\cite{Oti}) exhibit distinct NVLink transfer patterns. 

\noindent \textbf{{Experimental setup and data collection.}} Similar to our application fingerprint attack, 
the background spy program profiles the values of the specific counters: \texttt{nvlink\_total\_da\-ta\_received}, as well as timing delays resulting from the victim application's contention on NVLink. 
Our evaluation of the attack was conducted on the public cloud machine (a GCP 2-V100 machine). We discovered that Blender does not support configurations with 4-V100 or 8-V100 on GCP, as each pair of GPUs requires an NVLink connection. However, as depicted in Fig.~\ref{fig:gcp_topo}, the GCP GPU machine's ring topology does not fit for 4 or 8 GPUs. 
For victim 3D characters, we select 50 fully rigged characters from the Blender Studio open movies~\cite{blender_character}. Each character is rendered within the exact same background scene with the same camera angle and resolution. We split the dataset into training and testing using an 80/20 ratio.

\noindent \textbf{{Feature engineering and classification.}} Similarly, we exploit a sliding window approach with a window size of 200 to extract the same 12 time-series features as our application fingerprint attack.
We then utilize KNN, XGBoost, and LightGBM models to infer which 3D character the victim renders. 
As shown in Table{~\ref{tb:3d_results}}, LightGBM obtains the highest F1 score (86.86\%) when relying solely on contention-based timing leakage, with XGBoost following closely at 85.30\%. When performance counters are included, both XGBoost and LightGBM see further improvements, reaching F1 scores of 90.11\% and 91.56\%, respectively.

\begin{table}[tbh]
\caption{3D graphics character fingerprint performance: F1 (\%), Precision (\%), and Recall (\%) on GCP (2 GPUs). \timeicon = Contention-based timing leakage, \faCalculator = Leaky performance counters.}
\centering
\begin{tabular}{cc|c|c|c|}
\cline{3-5}
\textbf{}                                                  & \textbf{} & \textbf{F1}    & \textbf{Prec} & \textbf{Rec} \\ \hline \hline
\multicolumn{1}{|c|}{\multirow{3}{*}{\textbf{\timeicon}}} & KNN       & 61.37          & 63.21         & 64.50        \\ \cline{2-5} 
\multicolumn{1}{|c|}{}                                     & XGBoost   & 85.30          & 87.70         & 85.50        \\ \cline{2-5} 
\multicolumn{1}{|c|}{}                                     & LightGBM  & \textbf{86.86} & 88.99         & 87.00        \\ \hline \hline
\multicolumn{1}{|c|}{\multirow{3}{*}{\textbf{\timecounter}}} & KNN       & 59.74          & 62.71         & 62.50        \\ \cline{2-5} 
\multicolumn{1}{|c|}{}                                     & XGBoost   & 90.11          & 93.10         & 90.50        \\ \cline{2-5} 
\multicolumn{1}{|c|}{}                                     & LightGBM  & \textbf{91.56} & 94.11         & 92.00        \\ \hline
\end{tabular}
\label{tb:3d_results}
\end{table}




\section{Cross-VM Attack}
\label{sec:cross-instance}
During reverse engineering exercises, we found a surprising observation: NVlink counters expose visible leakage across GPUs that do not communicate directly, provided that there is a link connecting them (this link is not in use).   This potentially enables GPUs belonging to different VM instances to observe leakage across instances.  In this section, we first demonstrate and characterize this leakage between two VM instances on GCP. We then implement the 3D rendering fingerprinting side channel attack across the two VM instances using this leakage.

\noindent \textbf{{Prerequisite for this attack.}} The prerequisite for this attack is co-location, meaning the attacker must reside on the same physical host as the target victim. Previous research has extensively demonstrated co-location attacks in cloud systems~\cite{ristenpart2009hey,fang2021repttack,zhao2024last,zhao2024everywhere}. Specifically, Zhao et al.~\cite{zhao2024last,zhao2024everywhere} presented lightweight and highly successful co-location attacks on Google Cloud Run. Based on these findings, we assume that the attacker has successfully co-located on the same physical GPU instance as the victim. The attacker then exploits NVLink leakages to exfiltrate sensitive information.

\begin{figure}[tbh]
    \centering
    \includegraphics[width=0.45\textwidth]{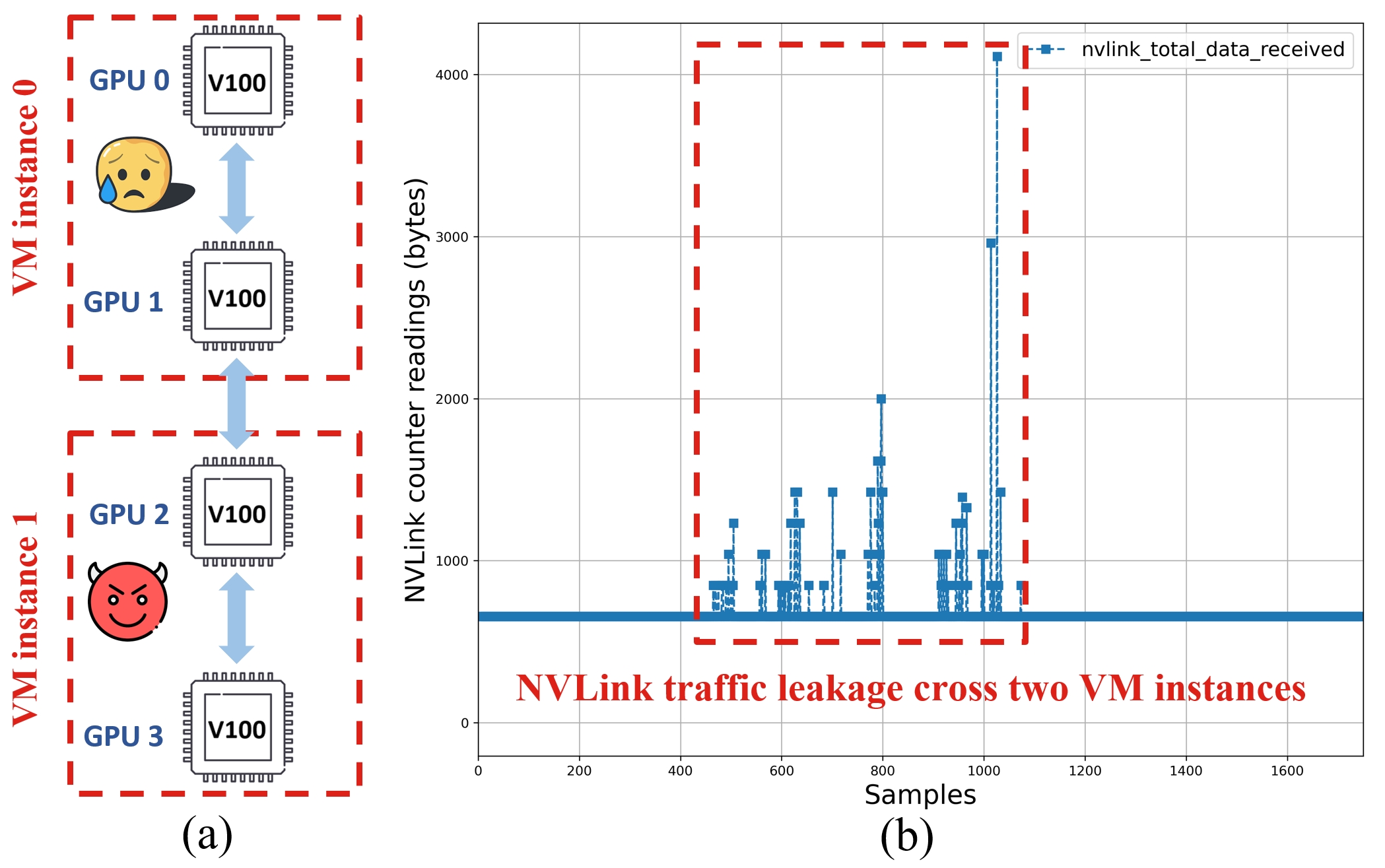}
    \caption{Cross-VM Leakage: (a) Instance topology:  Victim resides in VM Instance 0, 
    while the spy operates in Instance 1;
    (b) NVLink leakage:  Leakage across the two VM instances collected from GPU 2 in controlled by the spy's VM instance.}
    \label{fig:cross_instance}
\end{figure}

\noindent \textbf{{Experiment 4: Cross-VM leakage on the GCP.}} We assume the attacker is already co-located with the victim on the same physical host. To create co-located VMs, we first obtain a 4-V100 GPU instance on GCP, then we create two VMs within it using KVM (Ubuntu 20.04 host), assigning two GPUs to each VM.  As shown in Fig.~\ref{fig:cross_instance} (a), the victim's VM instance (Instance 0) includes GPUs 0 and 1, while the spy's VM instance (Instance 1) contains GPUs 2 and 3. NVLink is used as the interconnect in this topology. All experiments are conducted on the GCP Compute Engine in the us-west1-a region. All experiments were conducted within controlled environments, ensuring our research activities did not impact external public users.

In this experiment, the victim uses Blender to render 2D images from 3D holograms. The 3D object data is transmitted between GPU 0 and GPU 1 via NVLink. Simultaneously, the spy in another VM instance monitors the NvLink counter (\texttt{nvlink\_total\_data\_received}) on GPU 2. As shown in Table~\ref{tb:experiment_platform}, each V100 GPU is equipped with six NVLink slots, three of which are connected to peer GPUs. This allows GPU 2 in the spy's VM instance to monitor NVLink traffic patterns on the link between GPU 1 and GPU 2.

We observe that when two VM instances are connected, the spy can track NVLink traffic patterns from another VM instance by observing performance counters. Figure~\ref{fig:cross_instance} (b) depicts the NVLink traffic leakages collected across the two VM instances.  The signal measured on GPU 2, does not reflect the full traffic between GPU 1 and GPU 0, but it is correlated to it, providing consistent patterns for the different Blender characters.  This pattern can be observed as the sequence of spikes in the middle are caused by the program execution occurring within the victim's VM instance 0.




\begin{tcolorbox}[colback=gray!10,colframe=black,boxrule=0.5pt,arc=2mm,outer arc=2mm,]
\textit{\textbf{Observation 6}: When two VM instances are connected via NVLink, NVLink counters leak traffic patterns across VM instances even though this traffic does not occur on the physical link connecting the instances. }
\end{tcolorbox}

\noindent \textbf{{Attack 3: identifying rendered 3D characters across VMs.}} In this attack, we investigate whether the attacker can identify the 3D character the victim is rendering in a cross-VM scenario. Similar to our second attack, the background spy program monitors specific performance counters, including \texttt{nvlink\_total\_data\_received}, along with timing delays. The victim renders 50 fully rigged 3D characters from the Blender Studio open movies. Each character is rendered in the same background scene, using the same camera angle and resolution. The dataset is split into training and testing sets using an 80/20 ratio.

\begin{figure}[tbh]
    \centering
    \includegraphics[width=0.45\textwidth]{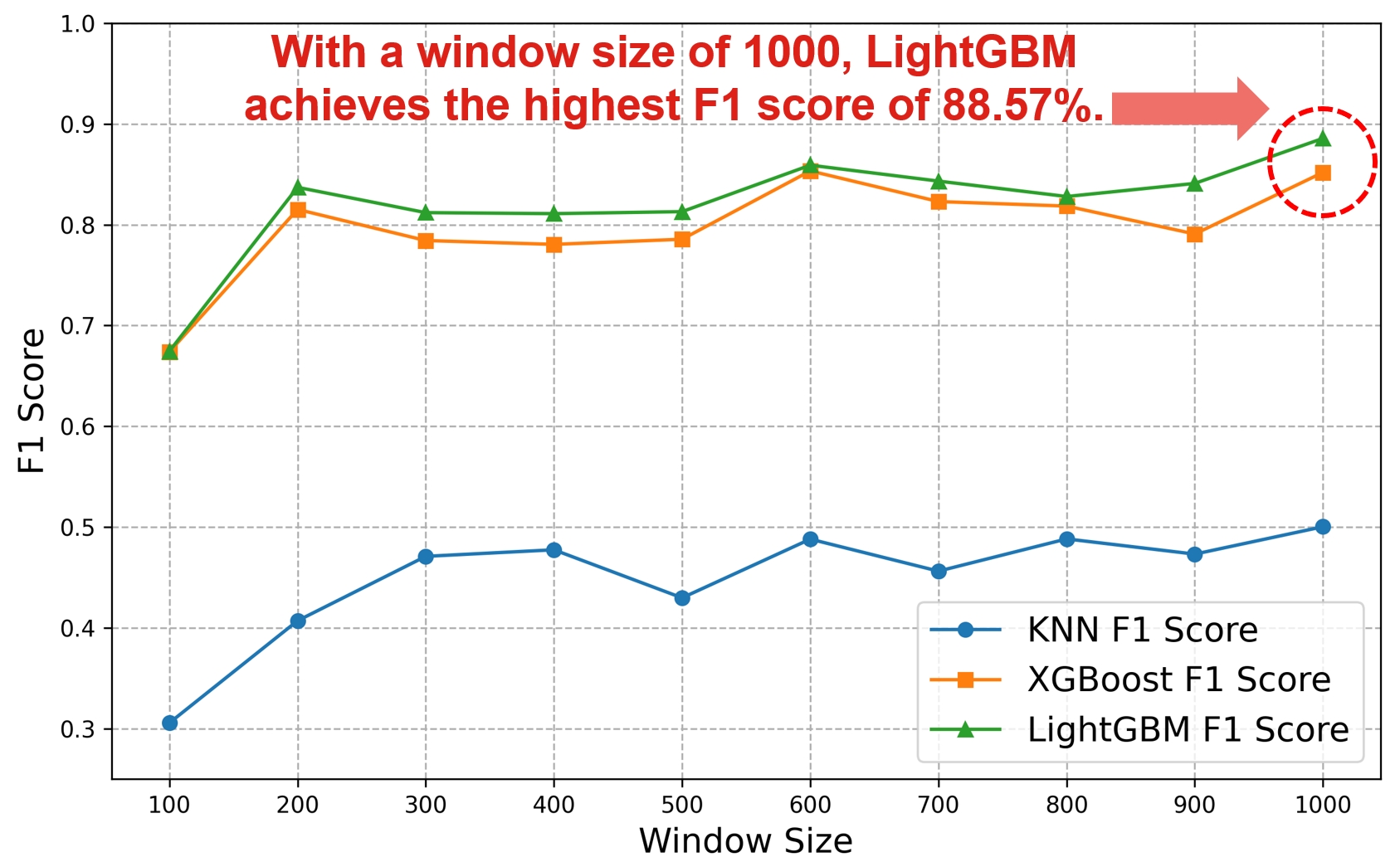}
    \caption{F1 scores with different window sizes.}
    \label{fig:cross_instance_results}
\end{figure}

\noindent \textbf{{Feature engineering and classification.}} Similarly, we use a sliding window approach with varied window sizes ranging from 100 to 1000 to extract the same 12 time-series features as in our application fingerprint attack. We then apply KNN, XGBoost, and LightGBM models (using the same hyperparameter settings as in Attack 2) to infer which 3D character the victim is rendering. Figure~\ref{fig:cross_instance_results} presents the performance (F1 scores) of the three classifiers for different window sizes.
We observe that with a window size of 1000, the LightGBM achieves the best performance, with an F1 score of 88.57\%, a precision of 91.16\%, and a recall of 88.50\%.

\section{Potential Mitigations} 
\label{sec:mitigation}

We discuss three potential classes of mitigations: (1) managing access to leaky counters, (2) restricting access to high-resolution clock timing instructions, and (3) detecting abnormal NVLink monitoring and/or contention.

\begin{figure}[tbh]
    \centering
    \includegraphics[width=0.45\textwidth]{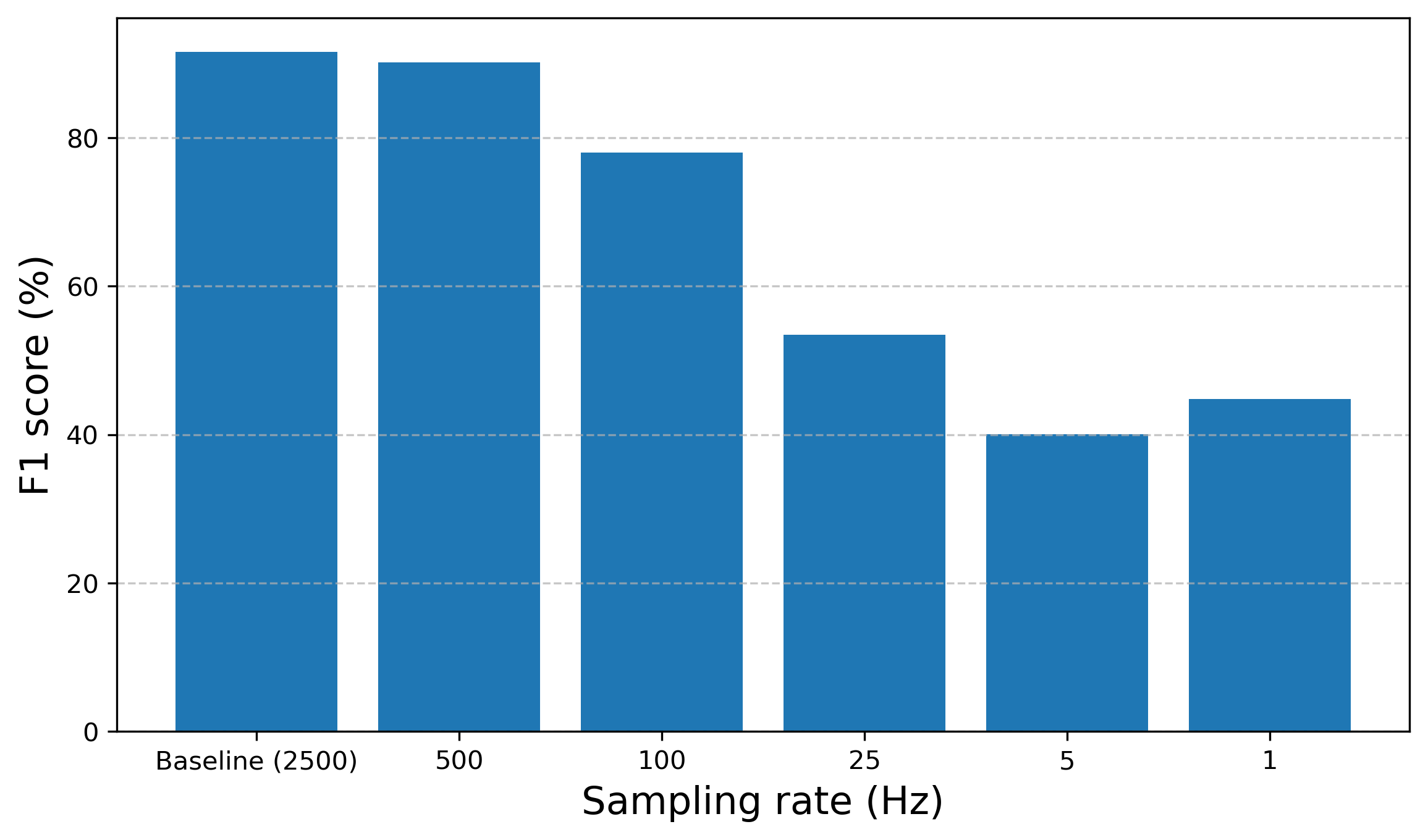}
    \caption{Accuracy with reduced sampling rate.}
    \label{fig:defense_subsampling}
\end{figure}

\noindent \textbf{{Managing access to counters.}} Reducing the sampling rate of performance counters may diminish the effectiveness of the spy application~\cite{naghibijouybari2018rendered, zhang2023s}. However, this approach could negatively impact legitimate users, as the functionality of these counters may not operate correctly. Fig.~\ref{fig:defense_subsampling} presents the impact of reducing the sampling rate on 3D character classification. Although the sampling rate is reduced to 1 Hz, the attacker can still gain information about the 3D character with an F1 score exceeding 40\%, significantly outperforming a random guess (2\%). We conjecture that even when the sampling rate is restricted to low levels, the leaky counters continue to retain valuable signals that assist attackers in inferring the activities of other users.

Completely blocking access to NVLink counters appears to be a viable defense against \textit{NVBleed}. However, our findings indicate that \textit{NVBleed} remains feasible even when attackers solely exploit timing characteristics to estimate communication behavior (e.g., Equation~\ref{equation:synthetic}). 
Using these estimated values, we retrained the models for a 3D character fingerprint attack: we were able to accurately identify the correct 3D character, achieving an F1 Score of over 83\% with the LightGBM model even when performance counters are not accessible.






\noindent \textbf{{Restricting access to high-resolution clock instructions.}} Our attacks utilize the high-resolution clock instruction, \textit{RDTSCP}, to measure the victim’s contention on a shared NVLink. Completely blocking access to these clock instructions could deter our attacks; however, legitimate users may also require such instructions to evaluate the performance of their programs.
On the other hand, even if specific clock instructions are prohibited, attackers can still utilize alternative low-requirement clock instructions (\eg Timer interrupts~\cite{schwarz2018keydrown}) or create a synthetic timer, as discussed in~\cite{dutta2023spy}.

\noindent \textbf{{Detecting abnormal NVLink monitoring and/or contention.}} \textit{NVBleed} operates by creating contention on shared NVLink interconnects and continuously probing the NVLink. A potential defense involves monitoring for such suspicious NVLink querying behaviors. 
GPUGuard~\cite{xu2019gpuguard} proposes a contention detection and dynamic partitioning framework designed to defend against covert and side channels in GPUs. However, their method primarily safeguards applications on a single GPU and is incompatible with multi-GPU systems. Additionally, the current hardware of NVLink does not support their methods of dynamic partitioning. 

\section{Related Work}
\label{sec:relatedwork}


\noindent \textbf{{Covert and Side-attacks in GPUs.}} Covert and side-channel attacks have been extensively studied in discrete GPUs. 
Naghibijouybari et al.~\cite{naghibijouybari2017constructing} introduced a contention-based covert channel across various GPU resources. Subsequent work conducted multiple side-channel attacks on both graphics and computational GPU workloads by monitoring GPU performance counters~\cite{naghibijouybari2018rendered,wei2020leaky}. 
Ahn et al.~\cite{ahn2021network} proposed a timing covert channel for GPUs that exploits the shared, on-chip interconnect channels. Additionally, Nayak et al.~\cite{nayak2021mis} developed a covert channel leveraging the GPU's shared last-level translation lookaside buffer (TLB). 

Recent studies also have highlighted several covert and side-channel attacks on integrated GPUs. 
Dutta et al.~\cite{dutta2021leaky} demonstrated covert and side-channel attacks between CPUs and GPUs through the shared Last Level Cache (LLC) and ring bus in integrated CPU-GPU systems. 
Yang et al.~\cite{yangSmartPhoneGPU} present a side-channel attack on mobile GPUs to eavesdrop on user credentials via GPU performance counters. Almusaddar et al.~\cite{almusaddar2023exploiting} exploited observable slowdown in shared read and write buffers within the memory controller and constructed a cross-processor covert channel in integrated CPU-GPU systems.   Wang et al.~\cite{wang2024gpu} identified a side-channel leakage during the graphical data compression process on integrated GPUs. 


\noindent \textbf{{Interconnect contention based attacks.}} Our attacks can also be classified as interconnect contention-based attacks. Previous research has shown that such leakages can be exploited to create covert and side channels. Tan et al.~\cite{tan2021invisible} introduced PCIe congestion side-channel attacks that span GPUs, Network Interface Cards (NICs), and Solid-State Drives (SSDs).
Similar contention-based attacks have been executed on other hardware architectures, such as the Last Level Cache (LLC)~\cite{liu2015last}, CPU ring bus~\cite{paccagnella2021lord}, the Host-GPU PCIe bus~\cite{side2022lockeddown}, and the CPU Mesh~\cite{wan2022meshup, dai2022don}. These prior attacks share a high-level observation with our attack: traffic from different sources on shared interconnects induces measurable contention. However, beyond this high-level similarity, our attack differs in several key aspects. 
Firstly, since we target multi-GPU systems, the threat model and attacker co-location options are significantly different. Prior attacks primarily focus on interconnects within a single CPU or GPU, requiring the attacker to co-locate with the victim. In contrast, our attack exploits inter-GPU communication, differing from traditional co-location, as the attacker can reside on a different GPU or even in a separate VM instance. 
Secondly, prior attacks typically exploit interconnects carrying cache traffic, memory transactions, or I/O operations, whereas our attack targets inter-GPU traffic. This distinction influences both the type of leakage exposed and the methodology for constructing end-to-end attacks. Beyond intra-VM attacks, our work demonstrates a cross-VM attack on a public cloud provider (GCP) and identifies a new leakage. 

The only prior work on microarchitectural attacks in multi-GPU systems is by Dutta et al.{~\cite{dutta2023spy}}, which employs prime-and-probe techniques to create cache contention via pinned memory pages. However, our attack differs in several key ways: (1). Communication-based leakage: We observe communication behavior rather than memory access patterns, revealing complementary leakage that can be combined for more powerful attacks. (2). No co-location requirement: Our attack does not require co-location, allowing an attacker to be on a remote GPU (even in a different VM instance) while still monitoring communication traffic between GPUs. By demonstrating the first attack on communication across multi-GPU interconnects, we expand the scope of contention-based side-channel research. This work highlights new security risks in multi-GPU systems and extends the side-channel threat model beyond traditional cache and memory-based attacks.


\noindent \textbf{Cross-VM attacks.} Groups of works have demonstrated the cross-VM attack on cloud systems. Ristenpart et al.~\cite{ristenpart2009hey} conducted the first study on co-residence detection in commercial cloud servers. Fang et al.~\cite{fang2021repttack} showed how attackers could manipulate cloud schedulers to achieve high co-location rates with target victims. Recently, Zhao et al.~\cite{zhao2024everywhere, zhao2024last} conducted co-location attacks in public cloud systems and successfully implemented an end-to-end LLC Prime+Probe side-channel attack on Google Cloud Run.

\section{Concluding Remarks}
\label{sec:conclusion}
In this paper, we demonstrate covert and side-channel attacks on multi-GPU interconnects. Through reverse engineering, we identify the communication patterns of two generations of NVLink. We develop two main classes of leakage vectors within NVLink: contention-based timing delays and leaky NVLink counters. Additionally, we demonstrate five specific attacks: two covert channels, two end-to-end side-channel attacks across multiple GPUs, and one side-channel attack across VMs. We believe the vulnerability of communication-based leakage might exist in other interconnected systems including other multi-GPU (and even multi-accelerator) systems, with different interconnects and potential leakage sources.

\bibliographystyle{plain}
\bibliography{main}





\end{document}